\documentclass[pdflatex,sn-mathphys-num]{sn-jnl}% Math and Physical Sciences Numbered Reference Style 
\usepackage{graphicx}%
\usepackage{multirow}%
\usepackage{amsmath,amssymb,amsfonts}%
\usepackage{amsthm}%
\usepackage{mathrsfs}%
\usepackage[title]{appendix}%
\usepackage{xcolor}%
\usepackage{textcomp}%
\usepackage{manyfoot}%
\usepackage{booktabs}%
\usepackage{algorithm}%
\usepackage{algorithmicx}%
\usepackage{algpseudocode}%
\usepackage{listings}%
\usepackage{tabularx}%
\usepackage{makecell}%
\theoremstyle{thmstyleone}%
\theoremstyle{thmstyletwo}%

\theoremstyle{thmstylethree}%

\begin{document}

\title[Article Title]{OAH-Net: A Deep Neural Network for Hologram Reconstruction of Off-axis Digital Holographic Microscope}


\author*[1]{\fnm{Wei} \sur{Liu}}\email{liuwei@bii.a-star.edu.sg}
\equalcont{These authors contributed equally to this work.}

\author[2,3]{\fnm{Kerem}\sur{Delikoyun}} \email{kerem.delikoyun@tum-create.edu.sg}
\equalcont{These authors contributed equally to this work.}

\author[2, 3]{\fnm{Qianyu}\sur{Chen}} \email{qianyu.chen@tum-create.edu.sg}
\author[3]{\fnm{Alperen}\sur{Yildiz}} \email{alperen.yildiz@tum-create.edu.sg}
\author[3]{\fnm{Si Ko}\sur{Myo}} \email{siko.myo@tum-create.edu.sg}

\author[4, 5]{\fnm{Win Sen}\sur{Kuan}} \email{win\_sen\_kuan@nuhs.edu.sg}
\author[4, 6]{\fnm{John Tshon Yit}\sur{Soong}} \email{John\_soong@nus.edu.sg}
\author[6]{\fnm{Matthew Edward}\sur{Cove}} \email{mdcmec@nus.edu.sg}
\author[2, 3]{\fnm{Oliver}\sur{Hayden}} \email{oliver.hayden@tum.de}
\author*[1, 7, 8, 9, 10]{\fnm{Hweekuan}\sur{Lee}} \email{leehk@bii.a-star.edu.sg}

\affil*[1]{\orgdiv{Bioinformatics Institute}, \orgname{Agency for Science, Technology and Research}, \orgaddress{\street{30 Biopilis Street}, \postcode{138671}, \country{Singapore}}}

\affil[2]{\orgdiv{School of Computation, Information and Technology}, \orgname{Technical University of Munich}, \orgaddress{\street{Arcisstr. 21}, \postcode{80333}, \city{Munich}, \country{Germany}}}

\affil[3]{\orgname{TUMCREATE}, \orgaddress{\street{1 Create Way}, \postcode{138602}, \country{Singapore}}}

\affil[4]{\orgdiv{Yong Loo Lin School of Medicine}, \orgname{National University of Singapore}, \orgaddress{\street{10 Medical Dr}, \postcode{117597}, \country{Singapore}}}

\affil[5]{\orgdiv{Emergency Medicine Department}, \orgname{National University Hospital}, \orgaddress{5 Lower Kent Ridge Road}, \postcode{119074}, \country{Singapore}}

\affil[6]{\orgdiv{Department of Medicine}, \orgname{National University Hospital}, \orgaddress{\street{5 Lower Kent Ridge Road}, \postcode{119074}, \country{Singapore}}}

\affil[7]{\orgdiv{Centre for Frontier AI Research}, \orgname{Agency for Science, Technology and Research}, \orgaddress{\street{1 Fusionopolis Way}, \postcode{138671}, \country{Singapore}}}

\affil[8]{\orgdiv{International Research Laboratory on Artificial Intelligence}, \orgname{Agency for Science, Technology and Research}, \orgaddress{\street{1 Fusionopolis Way}, \postcode{138671}, \country{Singapore}}}

\affil[9]{\orgdiv{School of Biological Sciences}, \orgname{Nanyang Technological University}, \orgaddress{\street{60 Nanyang Dr}, \postcode{639798}, \country{Singapore}}}

\affil[10]{\orgdiv{School of Computing}, \orgname{National University of Singapore}, \orgaddress{\street{13 Computing Dr}, \postcode{117417}, \country{Singapore}}}


\abstract{Off-axis digital holographic microscopy is a high-throughput, label-free imaging technology that provides three-dimensional, high-resolution information about samples, particularly useful in large-scale cellular imaging. However, the hologram reconstruction process poses a significant bottleneck for timely data analysis. To address this challenge, we propose a novel reconstruction approach that integrates deep learning with the physical principles of off-axis holography. We initialized part of the network weights based on the physical principle and then fine-tuned them via weakly supersized learning. Our off-axis hologram network (OAH-Net) retrieves phase and amplitude images with errors that fall within the measurement error range attributable to hardware, and its reconstruction speed significantly surpasses the microscope's acquisition rate. Crucially, OAH-Net demonstrates remarkable external generalization capabilities on unseen samples with distinct patterns and can be seamlessly integrated with other models for downstream tasks to achieve end-to-end real-time hologram analysis. This capability further expands off-axis holography's applications in both biological and medical studies.}


\maketitle

\section{Introduction}\label{sec1}

Digital holographic microscopy (DHM) is emerging as an innovative imaging modality in computational microscopy. It provides high-resolution, quantitative, and three-dimensional information about samples without labelling. These unique features make DHM a promising technique for imaging living cells, as it captures intracellular structures while preserving cells in their natural state, which could be used for more precise analysis \cite{charriere2006living, doi:10.1073/pnas.191361398, doi:10.1073/pnas.0806100105, Kemper2019}.

DHM records the interference pattern between the object and the reference beams, which is then reconstructed using algorithms to retrieve the wave of the object beam in terms of phase and amplitude \cite{off-axis_recontruction_review, chen2022fourier, huang2023self}. Holography can be classified into two main types based on beam alignment: inline holography and off-axis holography \cite{KOCH2010460, LATYCHEVSKAIA2010472}. For inline holography, the reference beam is parallel to the object beam. Although the setup is relatively simple, the hologram reconstruction requires multiple exposures of the same sample at varying sample-to-sensor distances. For off-axis holography, the reference beam is slightly titled to form a small angle with the object beam, resulting in spatial separation of the hologram in the frequency domain and subsequently facilitating reconstruction. Off-axis holography requires only one exposure per sample for reconstruction, making it an ideal technique for imaging nonstatic samples and high-throughput applications. For example, off-axis DHM has been used for blood-based cellular diagnostics by processing blood samples through a specialized microfluidic channel and imaging blood cells at high speed \cite{Ugele2018, Klenk2023}.

To fully leverage DHM's high-throughput imaging capabilities, high-speed analysis approaches are imperative to process a vast number of images in a timely manner. For complete blood cell counting, DHM could record more than 10,000 frames in 1.5 minutes. In our current pipeline \cite{Klenk2023}, it is infeasible to realize a typical turnaround time (TAT) of 30 min for heamotology analysis (Figure. \ref{fig: 1}b). Deep-learning techniques have demonstrated dominant advantages in computer vision, and some state-of-the-art methods have achieved real-time analysis speeds \cite{yolov5, yolov8_ultralytics, DINO, SwinTransformer}. However, computation-intensive holographic reconstruction hinders the true potential of high-throughput clinical diagnostic applications of DHM, leveraged with deep learning based data analysis frameworks. Therefore, the speed of hologram reconstruction in DHM while retrieving optically consistent phase and amplitude images has become the major bottleneck for any clinical applications with desirable turnaround time and clinical utility.

Two broad strategies have been used to improve the speed and accuracy of deep off-axis hologram reconstruction. The first strategy is purely data-driven, treating hologram reconstruction as a typical image-to-image task to convert holograms into phase and amplitude images \cite{HRNet, Zhang:18,app122010656, Y-net}. A major drawback of this approach is the performance drop when dealing with unseen samples distinct from the training data. This is particularly problematic in clinical settings, where patient data varies substantially across different disease phenotypes, medical instrumentation, or user intervention. The second category focuses on filtering signals in the frequency domain, taking advantage of the physical principle of off-axis holography. Many manual-designed frequency filters have been proposed, including but not limited to regular shape filters \cite{Cuche:00, mann2005high}, Butterworth filters \cite{Butterworth_Filter}, binary adaptive filters \cite{WENG20142633}, weighted adaptive filters \cite{WeightedAdaptiveSpatialFilter}, and iterative thresholding filters \cite{he2016automated}. Existing filters have certain limitations; they are either too rough \cite{Cuche:00, mann2005high, Butterworth_Filter}, need manual selection of a sample-dependent threshold \cite{WENG20142633, WeightedAdaptiveSpatialFilter}, or involve an iterative process for thresholding \cite{he2016automated}. In addition to manual-designed filters, Xiao et al. proposed a deep-learning approach to generate sample-dependent filters using a U-Net model \cite{CNNFilter}. However, this method requires manual annotation for each sample in the frequency domain, which is time-consuming and requires high expertise, making it difficult for users to train the model with their own data or fine-tune the pre-trained model.

Here, we aimed to test a weakly supervised deep learning approach that integrates with the physical principles of off-axis holography. Using our own clinical samples, we hypothesised that this would overcome the bottleneck for real-time analysis of holograms. Operations on the frequency spectrum, including but not limited to filtration, are converted into matrix multiplications with trainable parameters, which are further fine-tuned within a deep-learning framework. The ground truth is automatically generated, eliminating the need for manual annotation and leveraging the large datasets provided by high-throughput DHM. Our model, named the Off-Axis Hologram Network (OAH-Net), seamlessly incorporates the physical principles of off-axis DHM into its architecture. As a result, OAH-Net demonstrates robust external generalization capabilities, even for samples with distinct patterns not seen during training. Furthermore, OAH-Net is designed as an end-to-end solution, requiring no preprocessing or post-processing steps, such as phase unwrapping.

Another advantage of our model is that the hologram only passes through the model once, unlike many approaches that require an iterative process for reconstruction \cite{he2016automated, app122010656}. This one-time processing ensures that all the holograms within a batch can finish processing at the same time, taking full advantage of parallel computing. The inference speed of our method is less than 3 ms/frame, significantly less than the acquisition rate of the microscope camera (9.5 ms/frame), achieving real-time processing of high-resolution ($1536\times2048$ pixels) holograms. In addition, the reconstruction error is very low, falling within the range of measurement error inherent to the hardware. When tested in a subsequent task of object detection using YOLO, the performance of YOLO on the images reconstructed by OAH-Net is almost the same as that of the ground-truth method.

In this study, we present a novel holographic reconstruction and phase retrieval algorithm that leverages deep learning-based image processing to meet real-world clinical needs by outperforming various well-established algorithms with higher accuracy and external generalizability. Our contribution lies in demonstrating the clinical utility, cost-effectiveness, and potential of this advanced algorithm for practical diagnostic applications in healthcare settings.

\section{Result}
\subsection{Model architecture and training}
\begin{figure}[htbp]
    \centering
    \includegraphics[width=1\textwidth]{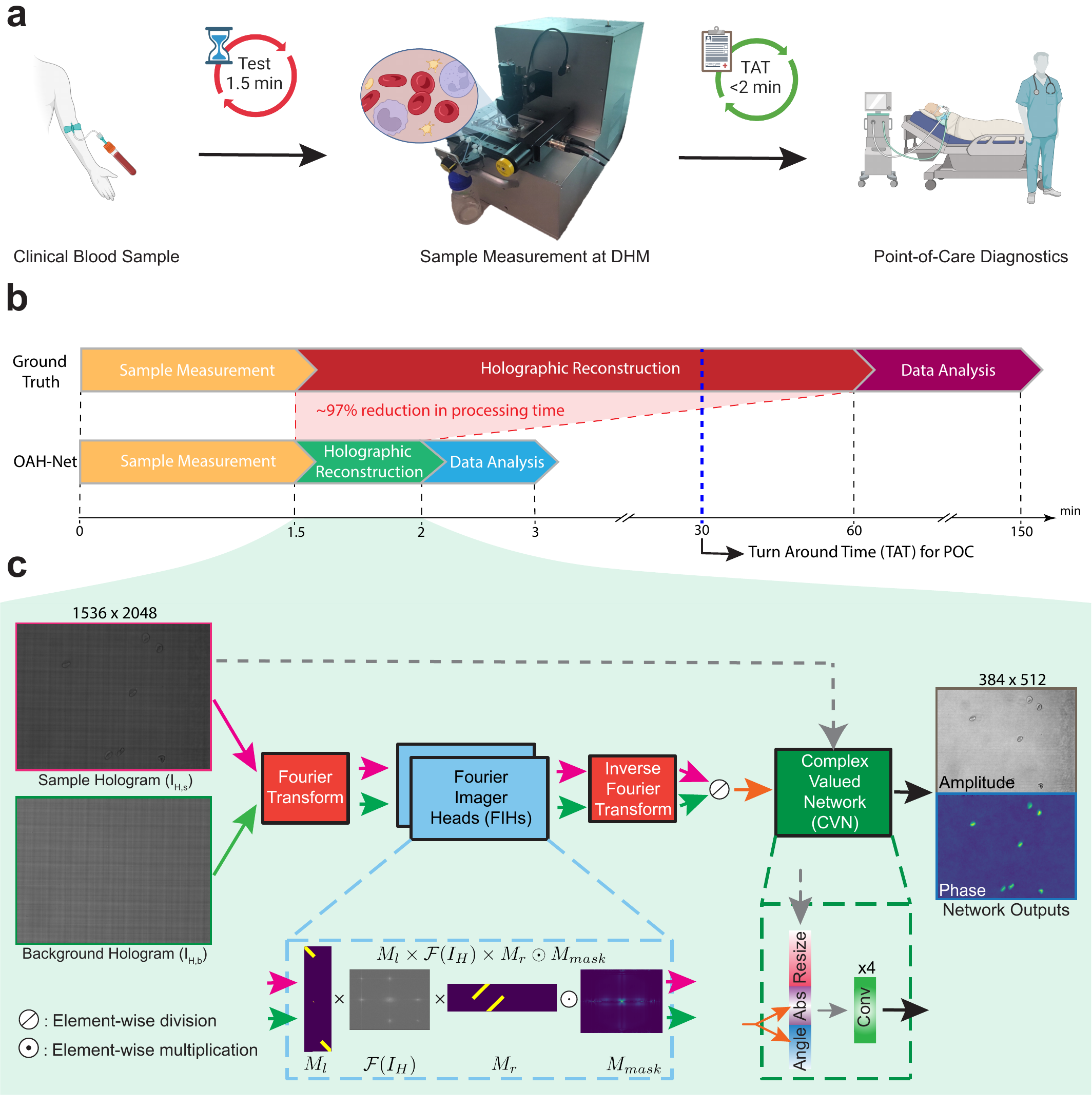}
    \caption {(a) Clinical workflow for point-of-care (POC) diagnostics using DHM, where clinical blood samples are screened at DHM. OAH-Net-based holographic reconstruction, integrating with an improved data analysis model, enables real-time analysis with a turnaround time (TAT) of under 5 minutes for a measurement typically involving 10,000 frames in 1.5 minutes. (b) Comparison of process breakdown between ground truth and OAH-Net-based workflows, showing OAH-Net dramatically accelerates image processing and achieves near real-time POC diagnosis. (c) The overall architecture of OAH-Net. OAH-Net consists of two main modules: Fourier Imager Heads (FIHs) and the complex-valued network (CVN). More details are provided in the Methods section. Although the size of the output images is one-quarter of the input image, there is no information loss. This is because the useful Fourier frequency occupies less than one-quarter of the overall spectrum.}
    \label{fig: 1}
\end{figure}

The architecture of OAH-Net is shown in Fig.\ref{fig: 1}c, comprising two main modules: Fourier Imager Heads (FIHs) and Complex Valued Network (CVN). FIHs are designed to filter out the undesired component of the hologram in the Fourier frequency domain and to correct the frequency shift caused by the titled reference beam. The CVN module converts the object beam wave in complex-valued form to amplitude and unwrapped phase. In some DHM devices, including the one used in our experiments, there is more than one reference beam, and the object beam wave can be derived from any of them individually. In such cases, the CVN module is also responsible for merging the multiple waves of the object beam into one. Further details are elucidated in the Methods Section.

The entire model is trained end-to-end in a supervised manner using blood cell samples. The input consists of holograms recorded in the presence of samples and a background hologram recorded without any sample. The target images, or ground truth, are the phase and amplitude images reconstructed from the sampled hologram via other methods. This study generated target images using Ovizio, an API software provided for our customized DHM with undisclosed technical details \cite{osone}.

\begin{figure}[htbp]
    \centering
    \includegraphics[width=1\textwidth]{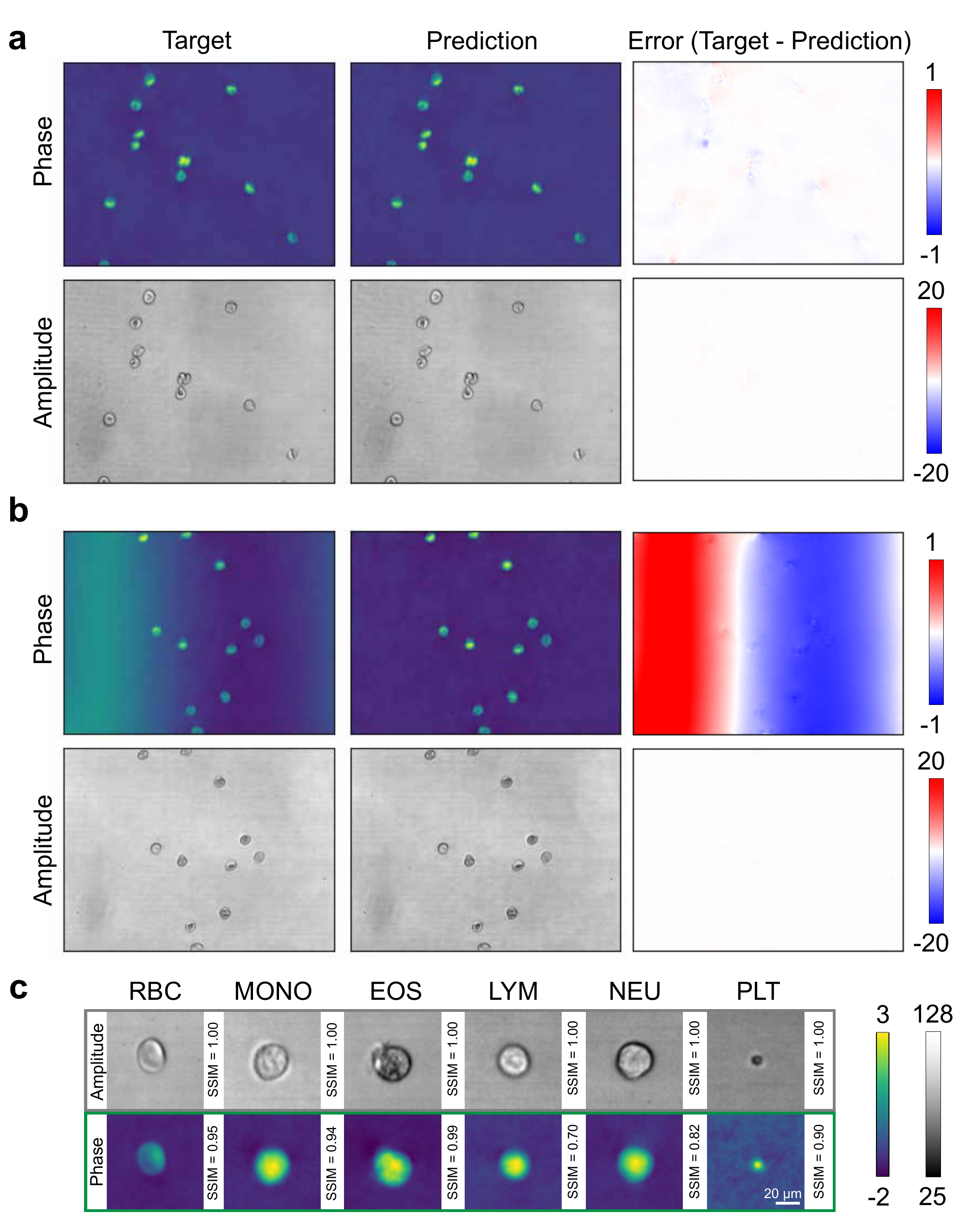}
    \caption{Comparison between the target images and the predicted images of OAH-Net in the test data. (a) A typical sample with a medium prediction error of phase. (b) The sample with the highest prediction error of phase. A low-frequency noise remains in the target phase image while it is successfully removed in the prediction of OAH-Net. (c) Blood cells of various types reconstructed via OAH-Net exhibit well-preserved intracellular structures, which can be utilized for downstream tasks. RBC, red blood cell; MONO, monocyte; EOS, eosinophil; LYM, Lymphocyte; NEU, neutrophil; PLT, platelet.}
    \label{fig: 2}
\end{figure}

\subsection{Model performance} We first assess the performance of the well-trained model in test data that recorded blood cell samples. Overall, OAH-Net's predictions are highly accurate. The error statistics are shown in Table\ref{tbl: benchmark}, and a typical example is presented in Fig.\ref{fig: 2}a. To compare with intrinsic measurement error, we recorded hologram videos of several static samples (as shown in Fig.\ref{fig: 4}a) and reconstructed the frames using Ovizio.api. The reconstructed images are not identical and exhibit minor fluctuations caused by hardware settings. The mean absolute error (MAE) averages for any two successive frames are $0.030\pm0.001$ for phase images and $0.677\pm0.003$ for amplitude images, higher than the prediction error of OAH-Net. Interestingly, in rare cases, low-frequency noise remains in the phase image reconstructed via Ovizio.api, as shown in Fig.\ref{fig: 2}b. In contrast, OAH-Net successfully removes the noise and generates a more homogeneous background, even though it was trained based on Ovizio.api.

\begin{figure}[htbp]
    \centering
    \includegraphics[width=1\textwidth]{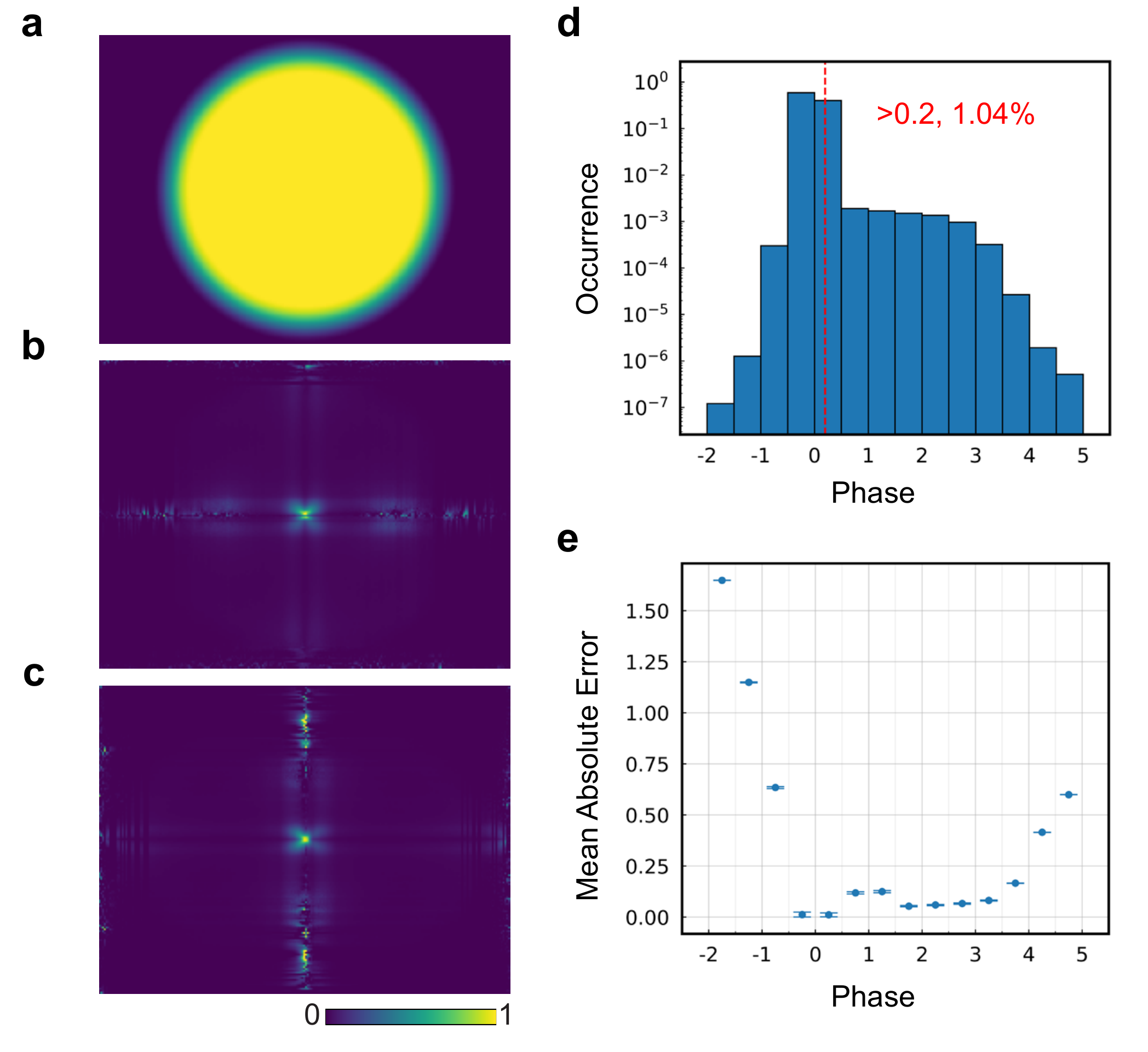}
    \caption{(a) The circular shape filter in the frequency domain \cite{Cuche:00}, which is also used as the initial weight of $M_{mask}$ of FIHs. (b-c) The fine-tuned $M_{mask}$ in both FIHs. All the filters are shifted to position the zero-frequency at the center for a better display. (d) The phase value distribution of all the ground truth images with blood cell samples. Pixels with phase values greater than $0.2$ constitute only $1.04\%$ of all pixels in the blood cell dataset. (e) Detailed MAE of OHA-Net predictions for each phase value range on the test dataset of blood cell sample.}
    \label{fig: 3}
\end{figure}

\begin{figure}[htbp]
    \centering
    \includegraphics[width=1\textwidth]{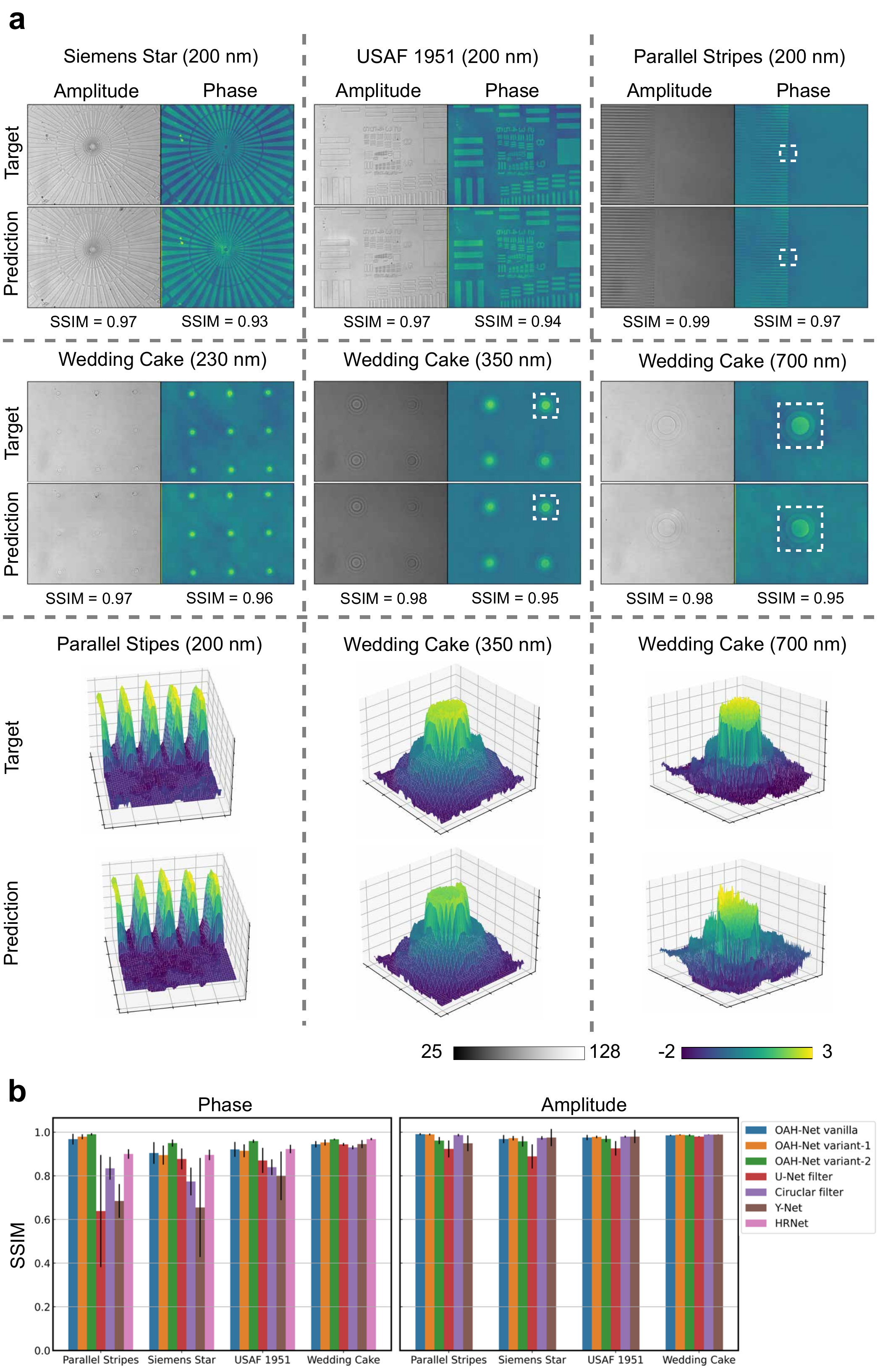}
    \caption{(a) External generalization of OAH-Net. The white dashed squares in the 2D images indicate the areas plotted in 3D. Despite being trained solely on blood cell samples, OAH-Net demonstrates high prediction accuracy when applied to static samples with distinct patterns. (b) Comparison of SSIM values for phase and amplitude reconstructions across different models. All wedding cake images were grouped for statistical analysis.}
    \label{fig: 4}
\end{figure}

The error measurement discussed above pertains to the entire image and is denoted as MAE\textsuperscript{1}. Practically, it is more important to accurately reconstruct regions with blood cells, defined as pixels with target phase values greater than $0.2$. We denote the MAE specifically for cell regions as MAE\textsuperscript{2}. MAE\textsuperscript{2} is slightly higher than MAE\textsuperscript{1}, possibly due to the imbalanced data distribution shown in Fig. \ref{fig: 3}d. The pixels in the cell region constitute only $1.04\%$ of all pixels in the blood cell dataset. Although the loss function is customized to encourage the model to focus more on the cell regions, the data imbalance issue cannot be solved entirely. Nevertheless, further analysis suggests that the error has negligible effects on downstream task performance. Fig.\ref{fig: 3}e shows that predicting pixels with negative target phase values is more error-prone. These pixels are usually part of the background, and their deviation between the target and prediction values plays a positive effect in making the background more homogeneous.

As shown in Eq. \ref{eq: FIH}, each FIH involves three trainable matrices: $M_l$, $M_r$, and $M_{mask}$, which are multiplied with the Fourier-transformed hologram $\mathcal{F}(I_H)$. $M_{mask}$ functions as a filter in the frequency domain, with the corresponding initial and fine-tuned weights shown in Fig. \ref{fig: 3}a-c. The fine-tuned $M_{mask}$ of both FIHs are distinct from the initial circular shape and exhibit interesting patterns that have not been reported before \cite{WeightedAdaptiveSpatialFilter}. The initial and fine-tuned weights of $M_l$ and $M_r$ are shown in Fig. \ref{fig: s3} and Fig. \ref{fig: s4}. Their initial weights consist only of 0 or 1, and are solely responsible for cropping and shifting the Fourier spectrum. After being fine-tuned as trainable parameters, $M_l$ and $M_r$ partially contributed to frequency filtration and improved model performance. During training, the weights of $M_l$, $M_r$, and $M_{mask}$ are not strictly restricted within the range of $[0, 1]$, but most of the fine-tuned values (approximately $99\%$) fall within this range. More detailed statistics are provided in the Supplementary.

\begin{table}[htbp]
    \centering
    \caption{The performance metrics of the benchmarks, OAH-Net, and OAH-Net variants on the test dataset of blood cell samples}
    \begin{tabular}{c c c c c  c c c c c}
    \hline
    \hline
    \multirow{2}{*}{\thead{\\Methods}}  & \multirow{2}{*}{\thead{Inference\\speed \\(ms/frame)}} & \multirow{2}{*}{\thead{Number of\\trainable\\parameters}} & \multicolumn{3}{c}{\thead{Phase (range:$-2\sim5$)}} & \multicolumn{3}{c}{\thead{Amplitude (range: $0\sim255$)}} \\
    \cmidrule(lr){4-6}\cmidrule(lr){7-9}
    & & & \thead{MAE\textsuperscript{1}} & \thead{MAE\textsuperscript{2}} & \thead{SSIM} & \thead{MAE\textsuperscript{1}} & \thead{MAE\textsuperscript{2}} & \thead{SSIM} \\
    \hline
    HRNet\cite{HRNet} & \makecell{28.54\\$\pm$0.54} & 2.8M & \makecell{0.052\\$\pm$ 2.71e-3} & \makecell{0.232\\ $\pm$3.31e-3} & \makecell{0.974\\$\pm$2.02e-3}& \textbf{---} & \textbf{---} & \textbf{---} \\
    Y-Net\cite{Y-net} & \makecell{122.95\\$\pm$0.09} & 14.7M & \makecell{0.104\\$\pm$1.37e-2} & \makecell{0.221\\$\pm$1.46e-2} & \makecell{0.957\\$\pm$6.50e-3} & \makecell{3.787\\$\pm$4.36e-1} & \makecell{5.161\\$\pm$3.72e-1} & \makecell{0.944\\$\pm$8.55e-3} \\
    \makecell{U-Net \\ filter \cite{CNNFilter}} & \makecell{19.11\\$\pm$0.57} & 4.0M & \makecell{0.028\\$\pm$5.19e-3} & \makecell{0.104\\$\pm$3.46e-2} & \makecell{0.993\\$\pm$2.00e-3} & \makecell{0.809\\$\pm$4.24e-2} & \makecell{3.75\\$\pm$3.02e-1} & \makecell{0.985\\$\pm$3.04e-3} \\
    \makecell{Circular \\ filter \cite{Cuche:00}} & \textbf{\makecell{2.65\\$\pm$0.30}} & 48.4K & \makecell{0.033\\$\pm$6.72e-3} & \makecell{0.122\\$\pm$5.39e-2} & \makecell{0.928\\$\pm$4.51e-3} & \makecell{0.368\\$\pm$5.27e-3} & \makecell{0.912\\$\pm$2.89e-3} & \makecell{0.995\\$\pm$3.77e-4} \\
    \makecell{OAH-Net \\ vanilla} & \textbf{\makecell{2.65\\$\pm$0.30}} & 3.7M & \textbf{\makecell{0.012\\$\pm$3.07e-4}} & \textbf{\makecell{0.049\\$\pm$9.29e-4}} & \textbf{\makecell{0.997\\$\pm$7.89e-5}} & \textbf{\makecell{0.372\\$\pm$1.65e-2}} & \textbf{\makecell{0.575\\$\pm$3.92e-2}} & \textbf{\makecell{0.995\\$\pm$2.19e-4}} \\
    \makecell{OAH-Net \\ variant-1} & \textbf{\makecell{2.65\\$\pm$0.30}} & 441K & \makecell{0.021\\$\pm$4.91e-4} & \makecell{0.085\\$\pm$1.41e-3} & \makecell{0.995\\$\pm$ 5.15e-5} & \makecell{0.454\\$\pm$7.53e-2} & \makecell{0.827\\$\pm$1.31e-2} &  \makecell{0.994\\$\pm$ 8.55e-4}\\
    \makecell{OAH-Net\\variant-2} & \makecell{$5.59$\\$\pm 0.52$}& 9.24M & \makecell{0.019\\$\pm$5.11e-4} & \makecell{0.066\\$\pm$2.32e-3} & \makecell{0.943\\$\pm$7.94e-3} & \makecell{1.898\\$\pm$2.18e-1} & \makecell{3.987\\$\pm$1.49e-1} & \makecell{0.986\\$\pm$9.25e-4}\\
    \hline
    \hline
    \end{tabular}
    \footnotetext{MAE\textsuperscript{1}, mean absolute error of the whole image; MAE\textsuperscript{2}, mean absolute error of the cell regions only; SSIM, structural similarity index measure.}
    \footnotetext{The inference speed is measured using an NVIDIA GeForce RTX 4090 GPU with a batch size of 1 for all methods. The inference speeds for the circular filter, OAH-Net, and OAH-Net variant-1 are identical as they share the same model architecture.}
    \label{tbl: benchmark}
\end{table}

\subsection{Benchmarks and model variations}
In Table \ref{tbl: benchmark}, we compared the performance of other selected methods, the OAH-Net and OAH-Net variants, in the blood cell sample test dataset. Methods requiring manual selection or iterative processing were excluded, as they are unsuitable for high-throughput analysis.

HRNet\cite{HRNet} and Y-Net\cite{Y-net} are standalone algorithms that adapt the purely data-driven strategy to directly convert holograms into phase images (HRNet) or both phase and amplitude images (Y-Net). Y-Net is based on U-Net \cite{U-Net}, while HRNet is based on ResNet \cite{ResNet}. We constructed these models as described in their respective reports, except for the last layers. The last layers of both models were revised to output images one-quarter the size of the input, consistent with our model. The loss function is the same as that used for our model, shown in Eq. \ref{eq: loss}. The weight of the amplitude error, $w_A$, was set to $0$ for HRNet, as there is no amplitude prediction. For Y-Net, $w_A$ was adjusted to $1$ to balance the errors in the phase and amplitude predictions. 

U-Net filter and the circular filter are two approaches to generate filters in the frequency domain. These are not standalone algorithms for hologram reconstruction. We integrate them, respectively, into the OAH-Net framework to replace $M_{mask}$, while fixing $M_l$ and $M_r$ as the initial value. The U-Net filter method generates a sample-dependent filter following the approach of Xiao et al.\cite{CNNFilter}. The original report trained the U-Net filter in a supervised manner by manually annotating frequency filters for each sample, which is time-consuming and potentially subjective. In this study, the U-Net filter is trained with the CVN module in the OAH-Net framework in an indirectly supervised manner. The circular filter is a classic regular-shape filter \cite{Cuche:00}, as shown in Fig. \ref{fig: 3}a, which could be converted to non-trainable $M_{mask}$. In such a case, all the trainable parameters of the OAH-Net framework are in the CVN module. Three different radii of the circular filter were tested, and the result with the best performance is shown in Table \ref{tbl: benchmark}.

As mentioned above, the matrices $M_l$ and $M_r$ in FIHs are primarily responsible for the cropping and shifting of the Fourier frequency. We build a variant of OAH-Net, denoted as OAH-Net variant-1 in Table \ref{tbl: benchmark}, to examine whether it is necessary to convert $M_l$ and $M_r$ into trainable parameters or training $M_{mask}$ alone is sufficient for frequency filtration. In variant-1, $M_l$ and $M_r$ are fixed with values of 0 or 1 only and are responsible solely for cropping and shifting Fourier frequencies, while the rest of the model is the same as the vanilla version. Table \ref{tbl: benchmark} shows that variant-1 performs slightly worse than the vanilla version of OAH-Net, suggesting that training $M_l$ and $M_r$ could help in frequency filtration. We also tested another variant of OAH-Net to explore whether a more complex CVN module could further improve model performance. In OAH-Net variant-2, the CVN adopts a U-Net architecture, following Xiao's report \cite{CNNFilter}. However, the results do not show any performance improvement.

All methods were trained three times with different initialization seeds, and the vanilla version of OAH-Net continuously outperforms other methods in all metrics.

\subsection{Model generalization ability}
To demonstrate the generalizability of OAH-Net, we tested the model, trained exclusively on blood cell samples, on unseen samples with distinct patterns. As shown in Fig.\ref{fig: 4}a, the model performs consistently well in various patterns, suggesting that it follows physical principles and without overfitting the training data. On the other hand, Fig. \ref{fig: 4}b shows a significant drop in phase SSIM for benchmark methods, particularly in parallel stripes and Siemens star sample. The overall good performance on Wedding cake sample is likely due to the large background area, with minimal phase or amplitude variance.

\subsection{Performance in downstream task}
In most cases, hologram reconstruction serves as a precursor to subsequent tasks such as classification and object detection for further sample analysis \cite{rohrl2023explainable}. To assess the broader utility of our model, we performed additional validation by integrating it with two object detection models, YOLOv5 and YOLOv8 \cite{yolov5, yolov8_ultralytics}. Our approach involved training and testing the YOLO models initially on target images, followed by repeating the process using images reconstructed by OAH-Net. Both YOLO models were trained using default settings without hyperparameter tuning. Performance comparison, detailed in Table \ref{tbl: yolo}, reveals no significant differences in object detection metrics between the two image sources. This outcome suggests that the accelerated hologram reconstruction achieved by our model does not entail any discernible performance trade-offs in subsequent object detection tasks. Furthermore, typical cells of various types reconstructed via OAH-Net are shown in Fig.\ref{fig: 2}c. The intracellular structures are well preserved and can be used for a more detailed analysis. Due to limited annotations, object detection models treat all white blood cells as the same class, disregarding their subclasses in this study.

\begin{table}[htbp]
    \centering
    \caption{Yolo models performance on target images and predicted images from the blood cell dataset}
    \begin{tabular}{c c | c c | c c | c c | c c}
    \hline
    \hline
    \multirow{3}{*}{\thead{\makecell{\\Image \\ source}}} & \multirow{3}{*}{\thead{\makecell{\\Object \\ type}}} & \multicolumn{4}{c}{\thead{YOLOv5}} & \multicolumn{4}{c}{\thead{YOLOv8}} \\
    \cmidrule(lr){3-6}\cmidrule(lr){7-10}
    & & \multicolumn{2}{c}{\thead{Amplitude}} & \multicolumn{2}{c}{\thead{Phase}} & \multicolumn{2}{c}{\thead{Amplitude}} & \multicolumn{2}{c}{\thead{Phase}} \\
    \cmidrule(lr){3-4}\cmidrule(lr){5-6} \cmidrule(lr){7-8}\cmidrule(lr){9-10}
     &  & \thead{mAP\textsubscript{50}} & \thead{mAP\textsubscript{95}} & \thead{mAP\textsubscript{50}} & \thead{mAP\textsubscript{95}} & \thead{mAP\textsubscript{50}} & \thead{mAP\textsubscript{95}} & \thead{mAP\textsubscript{50}} & \thead{mAP\textsubscript{95}}\\
    \hline
    \multirow{4}{*}{\makecell{Ovizio}}
    & RBC & \makecell{0.992\\$\pm$2.00e-2} & \makecell{0.922\\$\pm$2.00e-2}  & \makecell{0.993\\$\pm$4.71e-4} & \makecell{0.963\\$\pm$2.94e-3} & \makecell{0.993\\$\pm$2.08e-2} & \makecell{0.930\\$\pm$1.53e-2}& \makecell{0.993\\$\pm$4.71e-4} & \makecell{0.975\\$\pm$1.25e-3} \\
    & WBC & \makecell{0.993\\$\pm$1.53e-2} & \makecell{0.966\\$\pm$2.65e-2} & \makecell{0.995\\$\pm$4.71e-4} & \makecell{0.993\\$\pm$4.71e-4} & \makecell{0.995\\$\pm$1.00e-3} & \makecell{0.970\\$\pm$1.00e-3} & \makecell{0.995\\$\pm$4.71e-4} & \makecell{0.993\\$\pm$9.43e-3} \\
    & PLT & \makecell{0.956\\$\pm$4.58e-2} & \makecell{0.676\\$\pm$3.21e-2} & \makecell{0.983\\$\pm$1.00e-3} & \makecell{0.884\\$\pm$2.62e-3} & \makecell{0.959\\$\pm$5.77e-4} & \makecell{0.691\\$\pm$1.00e-3} & \makecell{0.984\\$\pm$1.70e-3} & \makecell{0.889\\$\pm$4.50e-3} \\
    \hline
    \multirow{4}{*}{\makecell{OAH-Net}}
    & RBC & \makecell{0.994\\$\pm$2.37e-3} & \makecell{0.923\\$\pm$4.51e-3} & \makecell{0.995\\$\pm$4.71e-4} & \makecell{0.974\\$\pm$4.71e-4} & \makecell{0.993\\$\pm$2.00e-3} & \makecell{0.932\\$\pm$2.00e-3} & \makecell{0.996\\$\pm$4.71e-4} & \makecell{0.966\\$\pm$8.16e-4} \\
    & WBC & \makecell{0.993\\$\pm$2.08e-3} & \makecell{0.972\\$\pm$6.03e-3} & \makecell{0.995\\$\pm$4.71e-4} & \makecell{0.990\\$\pm$1.70e-3} & \makecell{0.993\\$\pm$2.08e-3} & \makecell{0.846\\$\pm$1.00e-3} & \makecell{0.995\\$\pm$4.71e-4} & \makecell{0.990\\$\pm$4.71e-4} \\
    & PLT & \makecell{0.955\\$\pm$4.36e-3} & \makecell{0.684\\$\pm$4.16e-3} & \makecell{0.980\\$\pm$2.49e-3} & \makecell{0.842\\$\pm$9.43e-4} & \makecell{0.963\\$\pm$1.53e-3} & \makecell{0.697\\$\pm$1.00e-3} & \makecell{0.981\\$\pm$3.68e-3} & \makecell{0.842\\$\pm$4.50e-3} \\
    \hline
    \hline
    \end{tabular}
    \footnotetext{RBC, red blood cells; WBC, white blood cells; PLT, platelet. Overall, no significant performance difference exists when Yolo models were trained and tested on target images or predicted images by OAH-Net.}
    \label{tbl: yolo}
\end{table}

\section{Discussion}
We demonstrated an end-to-end network for off-axis hologram reconstruction to address the critical bottleneck in high-throughput DHM applications. By integrating deep learning models with the physical principles of off-axis holography, OAH-Net achieves high speed and accuracy in hologram reconstruction, outperforming state-of-the-art methods. Our model processes the original high-resolution hologram (1536$\times$2048 pixels) without downsizing or dividing it into smaller fields of view, thus preserving all detailed sample information for precise downstream analysis. Our tests showed that the reconstructed images using OAH-Net performed equally well in the object detection task of different types of blood cells using YOLOv5 and YOLOv8 compared to ground truth data. Meanwhile, streamlined computation allows OAH-Net to achieve an inference speed of less than 3 ms/frame, significantly faster than the acquisition rate of the microscope camera (9.5 ms/frame). This enables not only real-time hologram reconstruction but also potentially real-time end-to-end digital holographic microscopy-based diagnostic applications, including reconstruction and data analysis.

Compared to frameworks\cite{Ugele2018, Klenk2023}, OAH-Net reduces holographic reconstruction time by up to 97\% (excluding downstream data analysis), achieving a TAT of less than 5 minutes (Fig.\ref{fig: 1}b). In addition, OAH-Net addresses the challenge of high-throughput imaging storage by eliminating the need to store raw data. With real-time reconstruction, only the smaller reconstructed images (1/16 of the original size) are saved, significantly reducing storage requirements and costs. When integrated with downstream models to achieve real-time analysis, storage can be further optimized to save only frames with regions of interest or stored temporarily on memory,  allowing the use of more affordable hardware and enhancing the framework's practicality and scalability in clinical settings. Hence, OAH-Net not only enables real-time processing with minimal storage requirements but through hardware optimizations in the future (e.g., dedicated processing unit on FPGA) could further reduce the power consumption of DHM-based diagnostic devices, supporting the shift toward more sustainable healthcare solutions, addressing the growing demand for eco-friendly practices in the medical field.

OAH-Net demonstrated superior generalization capability across a broad spectrum of sample patterns distinct from the training data in zero-shot learning. This is a crucial advantage for DHM applications in the real world, where inference data distributions could significantly differ from training data distribution due to sample-related or user-related variations. However, it should be noted that the integrated physical principles are responsible only for frequency filtration. The phase unwrapping task is performed in the CVN module, which is purely data-driven. Hence, samples with phase values outside the training data range $[-2, 5]$ could be error-prone in hologram reconstruction, requiring fine-tuning of the model with the new data set. 

The target images used to train OAH-Net are automatically generated without manual annotation. In this study, they are derived via Ovizio.api software but could be generated using other numerical methods \cite{Verrier:11}. As demonstrated above, OAH-Net can generate more homogeneous images and is better at filtering low-frequency noise than the target image, indicating that it is robust to methods for ground truth generation.

To demonstrate OAH-Net's utility in downstream tasks, we trained YOLO models on phase and amplitude images generated by OAH-Net and Ovizio.api (Table \ref{tbl: yolo}). Results show that phase images generally outperform amplitude images in object detection, with OAH-Net's amplitude reconstructions slightly better than Ovizio.api’s. Although Ovizio.api serves as the ground truth, OAH-Net, trained on its output, achieved comparable performance. This indicates that OAH-Net effectively learns the same reconstruction patterns and generalizes well, especially in detecting small objects like platelets, where phase images consistently excel.

These enhancements make the OAH-Net an advantageous and practical solution for real-world applications, especially in clinical diagnostics, where speed and accuracy are paramount. We anticipate that the network architecture's inherent simplicity and demonstrated effectiveness will catalyze the further adoption of OAH-Net in quantitative phase imaging. Moreover, the potential applications of this approach extend beyond its current uses. It can be seamlessly integrated into various biomedical imaging modalities, especially where phase retrieval and direct processing into the frequency domain are essential. This adaptability and efficiency could mark a significant step forward in the field, promising to advance how we approach and utilize holographic imaging in medical diagnostics and biomedical research.

\section{Materials and methods}
\subsection{Digital holographic microscopy}
As shown in Fig.\ref{fig: s1}a, The customized DHM used in this study is provided by Ovizio Imaging Systems, Belgium, patented as "differential digital holographic microscopy" \cite{patent1, patent2}. The setup employs a 528 nm superluminescent light-emitting diode (SLED) from Osram for partially coherent Koehler illumination in transmission mode. It is equipped with a Nikon CFI LWD 40$\times$Cremove objective with a numerical aperture (NA) of 0.55.

Our data consists of de-identiﬁed hologram videos recording human whole blood samples flown inside microfluidic channel. All samples were taken from patients with written informed consent before blood draw in the Emergency Department and the Medical Intensive Care Unit of the National University Hospital of Singapore. Ethics approval was obtained from the National Healthcare Group Domain Specific Review Board (reference numbers 2021/00930 and 2021/01130). Blood samples were diluted with phosphate buffer saline (PBS, Sigma Aldrich) containing 0.05\% polyethylene oxide (PEO, Sigma Aldrich) to viscoelastically focus all blood cells flowing in a uniform plane, following the way of our previous studies \cite{Ugele2018, Klenk2023}. In total, 1672 videos were collected and randomly divided into 1038 for training, 331 for validation, and 303 for testing. Each video is approximately 1.5 minutes long with around 10,000 frames. However, due to storage limitations, only 200 frames were extracted from each video and used for this study.

Furthermore, we imaged a phase mask target (Quantitative Phase Target NIST Traceable, Benchmark Technologies, USA) to quantify our DHM setup's spatial resolution and phase retrieval capabilities. Different areas of this static sample with varying features and thicknesses are shown in Fig. \ref{fig: 4}a.

\subsection{Spatial separation of off-axis holograms in frequency domain}
In our DHM device setup, there is one object beam and two reference beams tilted along the perpendicular $x$ and $y$ axes (shown in Fig. \ref{fig: s1}b). The intensity distribution $I_H$ recorded by the CCD sensor is the interference pattern between the reference waves and the object wave:
\begin{align}
I_H &= (R_x + R_y + O)(R_x + R_y + O)^* \nonumber \\
&= R_xR_x^* + R_yR_y^* + OO^* + OR_x^* + R_xO^* + OR_y^* + R_yO^* + R_xR_y^* + R_x^*R_y \label{eq: I_H}
\end{align}
where $*$ represents the complex conjugate. Fig. \ref{fig: s1}c shows the terms in Eq. \ref{eq: I_H} can be effectively separated in the frequency domain, facilitating the exacting of $OR_x^*$ or $OR_y^*$ from $I_H$. The object wave $O$ can be reconstructed from either $OR_x^*$ or $OR_y^*$, with more details provided in the Supplementary.

\subsection{Deep learning framework for hologram reconstruction}
Our network architecture is shown in Fig. \ref{fig: 1}c. In our network, $OR^*$ separation and shifting in the Fourier frequency domain are performed by two matrix multiplications (denoted as $\times$) and one element-wise matrix multiplication (denoted as $\odot$) in the FIHs module:
\begin{equation}
F = M_l \times \mathcal{F}(I_H) \times M_r \odot M_{mask} \label{eq: FIH}
\end{equation}
where $\mathcal{F}$ is the Fourier transformation. $I_{H}$ is kept at high resolution with an original size of $1536\times2048$ pixels. The matrices $M_l$, $M_r$, and $M_{mask}$ are two-dimensional trainable matrices, with shapes $384\times1536$, $2048\times512$, and $384\times512$, respectively. $F$ is the output of FIH that is the corresponding Fourier frequency of $OR^*$. A set of $M_l$, $M_r$, and $M_{mask}$ is grouped as a Fourier Imager Head. In our study, two heads are used to output $OR_x^*$ and $OR_y^*$, respectively. The initialization of $M_l$, $M_r$, and $M_{mask}$ is based on the circular-shape filter, with further details provided in the Supplementary.

The output of FIHs, $OR_x^*$ and $OR_y^*$, is further demodulated to remove $R_x$ and $R_y$ via element-wise division to generate two object waves $O$ with minor differences. Subsequently, the two object waves $O$ are fed into the CVN module as input. The primary tasks of CVN include: 1) converting $O$ from the complex-value format into the phase-amplitude format, 2) merging multiple object waves $O$ into one, and 3) unwrapping the calculated phase. Task 1 requires no trainable parameters, while tasks 2 and 3 are implemented using CNN layers. Alternatively, the order can be switched, with the merging of multiple $O$ first in complex-value format using complex-valued CNN layers, followed by conversion to the phase-amplitude format. We compared these two strategies and found no significant performance differences. Hence, we opt for the first strategy to avoid complex-valued computation in the CNN layer, thereby enabling further acceleration of the model using techniques such as NVIDIA TensorRT. 

\subsection{Network training and testing}
We trained OAH-Net in a supervised manner using target amplitude ($A$) and phase ($\phi$) images generated by Ovizio.api. As shown in Fig. \ref{fig: 3}d, there is a data imbalance between the background pixels and the sample pixels. Hence, we use a weighted L1 loss function to guide the model's focus on areas of higher importance:
\begin{equation}
    L_{loss} = \frac{1}{n}\sum_{i=1}^n\sum_{j=1}^w\sum_{k=1}^h(|\hat{\phi}^i_{jk} - \phi^i_{jk}| + w_{A} \times |\hat{A}^i_{jk} - A^i_{jk}|) \times W(\hat{\phi}^i_{jk}, \phi^i_{jk}) \label{eq: loss}
\end{equation}
where $\hat{A}$ and $\hat{\phi}$ are the predicted amplitude and phase, respectively, $j$ and $k$ are the pixel coordinates, while $w$ and $h$ are the width and height of the image, respectively. $i$ indicate the $i$th sample and $n$ is the batch size. $w_A$ is the weight for amplitude errors, with a default value of $0.1$. $W$ is the pixel weight function:
\begin{equation*}
    W_{jk}(\hat{\phi}_{jk}, \phi_{jk}) = 40 \times \text{clamp}[ \text{max}(|\hat{\phi}_{jk}|, |\phi_{jk}|), 0, 0.05 ] + 20 \times \text{clamp}[\text{grad}(\phi_{jk}), 0, 0.1] + 1
\end{equation*}
where
\begin{equation*}
    \text{clamp}(x, a, b) = \max [a, \min(x, b)]
\end{equation*}
$\text{grad}(\phi_{jk})$ is the gradient of the target phase image derived by the Sobel filter \cite{Sobel} followed by a Gaussian blur.

All networks mentioned in this paper were trained using the Adam optimizer \cite{Adam} with a constant learning rate optimized by grid search. Training was halted when the validation loss showed no improvement for 200 successive epochs, and the model with the lowest validation loss was selected for further testing. The inference speed was measured with a batch size of 1 on a NVIDIA GeForce RTX 4090 GPU, after compiling the networks with the Pytorch JIT compiler \cite{Pytorch}. The structural similarity index (SSIM) was measured using the corresponding function in scikit-image \cite{scikit-image}, where the value range of the phase image was normalized from $(-2, 5)$ to $(0, 255)$, making it the same as amplitude images for easier comparison.

\backmatter

\bmhead{Supplementary information}

See the Appendix for supporting content. 

\bmhead{Acknowledgements}
This research is supported by the National Research Foundation, Prime Minister’s Office, Singapore under Intra-CREATE Thematic Grant (NRF2019-THE002-0008). The computational work for this article was partially performed on resources of the National Supercomputing Centre (NSCC), Singapore (https://www.nscc.sg). 

\bmhead{Conflict of interest}
W.L., K.D., Q.C., S.K.M., O.H., and H.L. are inventors of a pending patent related to this work filed by the Singapore Patent Office (10202401892U, filed on 26 June 2024). The authors declare that they have no other competing interests.

\bmhead{Data availability} Holograms and the reconstructed phase and amplitude via Ovizio for the static samples are available at \texttt{https://doi.org/10.6084/m9.figshare.27108547.v1}

\bmhead{Code availability}
Code of the study is available from the corresponding author on request. 

\bibliography{ref}


\begin{thebibliography}{38}
\ifx \bisbn   \undefined \def \bisbn  #1{ISBN #1}\fi
\ifx \binits  \undefined \def \binits#1{#1}\fi
\ifx \bauthor  \undefined \def \bauthor#1{#1}\fi
\ifx \batitle  \undefined \def \batitle#1{#1}\fi
\ifx \bjtitle  \undefined \def \bjtitle#1{#1}\fi
\ifx \bvolume  \undefined \def \bvolume#1{\textbf{#1}}\fi
\ifx \byear  \undefined \def \byear#1{#1}\fi
\ifx \bissue  \undefined \def \bissue#1{#1}\fi
\ifx \bfpage  \undefined \def \bfpage#1{#1}\fi
\ifx \blpage  \undefined \def \blpage #1{#1}\fi
\ifx \burl  \undefined \def \burl#1{\textsf{#1}}\fi
\ifx \doiurl  \undefined \def \doiurl#1{\url{https://doi.org/#1}}\fi
\ifx \betal  \undefined \def \betal{\textit{et al.}}\fi
\ifx \binstitute  \undefined \def \binstitute#1{#1}\fi
\ifx \binstitutionaled  \undefined \def \binstitutionaled#1{#1}\fi
\ifx \bctitle  \undefined \def \bctitle#1{#1}\fi
\ifx \beditor  \undefined \def \beditor#1{#1}\fi
\ifx \bpublisher  \undefined \def \bpublisher#1{#1}\fi
\ifx \bbtitle  \undefined \def \bbtitle#1{#1}\fi
\ifx \bedition  \undefined \def \bedition#1{#1}\fi
\ifx \bseriesno  \undefined \def \bseriesno#1{#1}\fi
\ifx \blocation  \undefined \def \blocation#1{#1}\fi
\ifx \bsertitle  \undefined \def \bsertitle#1{#1}\fi
\ifx \bsnm \undefined \def \bsnm#1{#1}\fi
\ifx \bsuffix \undefined \def \bsuffix#1{#1}\fi
\ifx \bparticle \undefined \def \bparticle#1{#1}\fi
\ifx \barticle \undefined \def \barticle#1{#1}\fi
\bibcommenthead
\ifx \bconfdate \undefined \def \bconfdate #1{#1}\fi
\ifx \botherref \undefined \def \botherref #1{#1}\fi
\ifx \url \undefined \def \url#1{\textsf{#1}}\fi
\ifx \bchapter \undefined \def \bchapter#1{#1}\fi
\ifx \bbook \undefined \def \bbook#1{#1}\fi
\ifx \bcomment \undefined \def \bcomment#1{#1}\fi
\ifx \oauthor \undefined \def \oauthor#1{#1}\fi
\ifx \citeauthoryear \undefined \def \citeauthoryear#1{#1}\fi
\ifx \endbibitem  \undefined \def \endbibitem {}\fi
\ifx \bconflocation  \undefined \def \bconflocation#1{#1}\fi
\ifx \arxivurl  \undefined \def \arxivurl#1{\textsf{#1}}\fi
\csname PreBibitemsHook\endcsname

%%% 1
\bibitem[\protect\citeauthoryear{Charri{\`e}re et~al.}{2006}]{charriere2006living}
\begin{barticle}
\bauthor{\bsnm{Charri{\`e}re}, \binits{F.}},
\bauthor{\bsnm{Pavillon}, \binits{N.}},
\bauthor{\bsnm{Colomb}, \binits{T.}},
\bauthor{\bsnm{Depeursinge}, \binits{C.}},
\bauthor{\bsnm{Heger}, \binits{T.J.}},
\bauthor{\bsnm{Mitchell}, \binits{E.A.D.}},
\bauthor{\bsnm{Marquet}, \binits{P.}},
\bauthor{\bsnm{Rappaz}, \binits{B.}}:
\batitle{Living specimen tomography by digital holographic microscopy: Morphometry of testate amoeba}.
\bjtitle{Optics Express}
\bvolume{14}(\bissue{16}),
\bfpage{7005}--\blpage{7013}
(\byear{2006})
\doiurl{10.1364/oe.14.007005}
\end{barticle}
\endbibitem

%%% 2
\bibitem[\protect\citeauthoryear{Xu et~al.}{2001}]{doi:10.1073/pnas.191361398}
\begin{barticle}
\bauthor{\bsnm{Xu}, \binits{W.}},
\bauthor{\bsnm{Jericho}, \binits{M.H.}},
\bauthor{\bsnm{Meinertzhagen}, \binits{I.A.}},
\bauthor{\bsnm{Kreuzer}, \binits{H.J.}}:
\batitle{Digital in-line holography for biological applications}.
\bjtitle{Proceedings of the National Academy of Sciences}
\bvolume{98}(\bissue{20}),
\bfpage{11301}--\blpage{11305}
(\byear{2001})
\doiurl{10.1073/pnas.191361398}
{\href{https://arxiv.org/abs/https://www.pnas.org/doi/pdf/10.1073/pnas.191361398}{{https://www.pnas.org/doi/pdf/10.1073/pnas.191361398}}}
\end{barticle}
\endbibitem

%%% 3
\bibitem[\protect\citeauthoryear{Park et~al.}{2008}]{doi:10.1073/pnas.0806100105}
\begin{barticle}
\bauthor{\bsnm{Park}, \binits{Y.}},
\bauthor{\bsnm{Diez-Silva}, \binits{M.}},
\bauthor{\bsnm{Popescu}, \binits{G.}},
\bauthor{\bsnm{Lykotrafitis}, \binits{G.}},
\bauthor{\bsnm{Choi}, \binits{W.}},
\bauthor{\bsnm{Feld}, \binits{M.S.}},
\bauthor{\bsnm{Suresh}, \binits{S.}}:
\batitle{Refractive index maps and membrane dynamics of human red blood cells parasitized by plasmodium falciparum}.
\bjtitle{Proceedings of the National Academy of Sciences of the United States of America}
\bvolume{105}(\bissue{37}),
\bfpage{13730}--\blpage{13735}
(\byear{2008})
\doiurl{10.1073/pnas.0806100105}
{\href{https://arxiv.org/abs/https://www.pnas.org/doi/pdf/10.1073/pnas.0806100105}{{https://www.pnas.org/doi/pdf/10.1073/pnas.0806100105}}}
\end{barticle}
\endbibitem

%%% 4
\bibitem[\protect\citeauthoryear{Kemper et~al.}{2019}]{Kemper2019}
\begin{bbook}
\bauthor{\bsnm{Kemper}, \binits{B.}},
\bauthor{\bsnm{Bauwens}, \binits{A.}},
\bauthor{\bsnm{Bettenworth}, \binits{D.}},
\bauthor{\bsnm{G{\"o}tte}, \binits{M.}},
\bauthor{\bsnm{Greve}, \binits{B.}},
\bauthor{\bsnm{Kastl}, \binits{L.}},
\bauthor{\bsnm{Ketelhut}, \binits{S.}},
\bauthor{\bsnm{Lenz}, \binits{P.}},
\bauthor{\bsnm{Mues}, \binits{S.}},
\bauthor{\bsnm{Schnekenburger}, \binits{J.}},
\bauthor{\bsnm{Vollmer}, \binits{A.}}:
In: \beditor{\bsnm{Wegener}, \binits{J.}} (ed.)
\bbtitle{Label-Free Quantitative In Vitro Live Cell Imaging with Digital Holographic Microscopy},
pp. \bfpage{219}--\blpage{272}.
\bpublisher{Springer},
\blocation{Cham}
(\byear{2019}).
\doiurl{10.1007/11663_2019_6}
\end{bbook}
\endbibitem

%%% 5
\bibitem[\protect\citeauthoryear{Verrier and Atlan}{2011}]{off-axis_recontruction_review}
\begin{barticle}
\bauthor{\bsnm{Verrier}, \binits{N.}},
\bauthor{\bsnm{Atlan}, \binits{M.}}:
\batitle{Off-axis digital hologram reconstruction: some practical considerations}.
\bjtitle{Appl. Opt.}
\bvolume{50}(\bissue{34}),
\bfpage{136}--\blpage{146}
(\byear{2011})
\doiurl{10.1364/AO.50.00H136}
\end{barticle}
\endbibitem

%%% 6
\bibitem[\protect\citeauthoryear{Chen et~al.}{2022}]{chen2022fourier}
\begin{barticle}
\bauthor{\bsnm{Chen}, \binits{H.}},
\bauthor{\bsnm{Huang}, \binits{L.}},
\bauthor{\bsnm{Liu}, \binits{T.}},
\bauthor{\bsnm{al.}}:
\batitle{Fourier imager network (fin): A deep neural network for hologram reconstruction with superior external generalization}.
\bjtitle{Light: Science \& Applications}
\bvolume{11}(\bissue{1}),
\bfpage{254}
(\byear{2022})
\doiurl{10.1038/s41377-022-00949-8}
\end{barticle}
\endbibitem

%%% 7
\bibitem[\protect\citeauthoryear{Huang et~al.}{2023}]{huang2023self}
\begin{barticle}
\bauthor{\bsnm{Huang}, \binits{L.}},
\bauthor{\bsnm{Chen}, \binits{H.}},
\bauthor{\bsnm{Liu}, \binits{T.}},
\bauthor{\bsnm{al.}}:
\batitle{Self-supervised learning of hologram reconstruction using physics consistency}.
\bjtitle{Nature Machine Intelligence}
\bvolume{5}(\bissue{12}),
\bfpage{895}--\blpage{907}
(\byear{2023})
\doiurl{10.1038/s42256-023-00704-7}
\end{barticle}
\endbibitem

%%% 8
\bibitem[\protect\citeauthoryear{Koch and Lubk}{2010}]{KOCH2010460}
\begin{barticle}
\bauthor{\bsnm{Koch}, \binits{C.T.}},
\bauthor{\bsnm{Lubk}, \binits{A.}}:
\batitle{Off-axis and inline electron holography: A quantitative comparison}.
\bjtitle{Ultramicroscopy}
\bvolume{110}(\bissue{5}),
\bfpage{460}--\blpage{471}
(\byear{2010})
\doiurl{10.1016/j.ultramic.2009.11.022} .
\bcomment{Hannes Lichte 65th Birthday}
\end{barticle}
\endbibitem

%%% 9
\bibitem[\protect\citeauthoryear{Latychevskaia et~al.}{2010}]{LATYCHEVSKAIA2010472}
\begin{barticle}
\bauthor{\bsnm{Latychevskaia}, \binits{T.}},
\bauthor{\bsnm{Formanek}, \binits{P.}},
\bauthor{\bsnm{Koch}, \binits{C.T.}},
\bauthor{\bsnm{Lubk}, \binits{A.}}:
\batitle{Off-axis and inline electron holography: Experimental comparison}.
\bjtitle{Ultramicroscopy}
\bvolume{110}(\bissue{5}),
\bfpage{472}--\blpage{482}
(\byear{2010})
\doiurl{10.1016/j.ultramic.2009.12.007} .
\bcomment{Hannes Lichte 65th Birthday}
\end{barticle}
\endbibitem

%%% 10
\bibitem[\protect\citeauthoryear{Ugele et~al.}{2018}]{Ugele2018}
\begin{barticle}
\bauthor{\bsnm{Ugele}, \binits{M.}},
\bauthor{\bsnm{Weniger}, \binits{M.}},
\bauthor{\bsnm{Stanzel}, \binits{M.}},
\bauthor{\bsnm{Bassler}, \binits{M.}},
\bauthor{\bsnm{Krause}, \binits{S.W.}},
\bauthor{\bsnm{Friedrich}, \binits{O.}},
\bauthor{\bsnm{Hayden}, \binits{O.}},
\bauthor{\bsnm{Richter}, \binits{L.}}:
\batitle{Label-free high-throughput leukemia detection by holographic microscopy}.
\bjtitle{Advanced Science}
\bvolume{5}(\bissue{12}),
\bfpage{1800761}
(\byear{2018})
\doiurl{10.1002/advs.201800761}
{\href{https://arxiv.org/abs/https://onlinelibrary.wiley.com/doi/pdf/10.1002/advs.201800761}{{https://onlinelibrary.wiley.com/doi/pdf/10.1002/advs.201800761}}}
\end{barticle}
\endbibitem

%%% 11
\bibitem[\protect\citeauthoryear{Klenk et~al.}{2023}]{Klenk2023}
\begin{barticle}
\bauthor{\bsnm{Klenk}, \binits{C.}},
\bauthor{\bsnm{Erber}, \binits{J.}},
\bauthor{\bsnm{Fresacher}, \binits{D.}},
\bauthor{\bsnm{Röhrl}, \binits{S.}},
\bauthor{\bsnm{Lengl}, \binits{M.}},
\bauthor{\bsnm{Heim}, \binits{D.}},
\bauthor{\bsnm{Irl}, \binits{H.}},
\bauthor{\bsnm{Schlegel}, \binits{M.}},
\bauthor{\bsnm{Haller}, \binits{B.}},
\bauthor{\bsnm{Lahmer}, \binits{T.}},
\bauthor{\bsnm{Diepold}, \binits{K.}},
\bauthor{\bsnm{Rasch}, \binits{S.}},
\bauthor{\bsnm{Hayden}, \binits{O.}}:
\batitle{Platelet aggregates detected using quantitative phase imaging associate with covid-19 severity}.
\bjtitle{Communications Medicine}
\bvolume{3}(\bissue{1}),
\bfpage{161}
(\byear{2023})
\doiurl{10.1038/s43856-023-00395-6}
\end{barticle}
\endbibitem

%%% 12
\bibitem[\protect\citeauthoryear{Jocher}{2020}]{yolov5}
\begin{botherref}
\oauthor{\bsnm{Jocher}, \binits{G.}}:
YOLOv5 by Ultralytics.
\doiurl{10.5281/zenodo.3908559} .
\url{https://github.com/ultralytics/yolov5}
\end{botherref}
\endbibitem

%%% 13
\bibitem[\protect\citeauthoryear{Jocher et~al.}{2023}]{yolov8_ultralytics}
\begin{botherref}
\oauthor{\bsnm{Jocher}, \binits{G.}},
\oauthor{\bsnm{Chaurasia}, \binits{A.}},
\oauthor{\bsnm{Qiu}, \binits{J.}}:
Ultralytics YOLOv8.
\url{https://github.com/ultralytics/ultralytics}
\end{botherref}
\endbibitem

%%% 14
\bibitem[\protect\citeauthoryear{Zhang et~al.}{2022}]{DINO}
\begin{botherref}
\oauthor{\bsnm{Zhang}, \binits{H.}},
\oauthor{\bsnm{Li}, \binits{F.}},
\oauthor{\bsnm{Liu}, \binits{S.}},
\oauthor{\bsnm{Zhang}, \binits{L.}},
\oauthor{\bsnm{Su}, \binits{H.}},
\oauthor{\bsnm{Zhu}, \binits{J.}},
\oauthor{\bsnm{Ni}, \binits{L.M.}},
\oauthor{\bsnm{Shum}, \binits{H.-Y.}}:
DINO: DETR with Improved DeNoising Anchor Boxes for End-to-End Object Detection
(2022).
\url{https://arxiv.org/abs/2203.03605}
\end{botherref}
\endbibitem

%%% 15
\bibitem[\protect\citeauthoryear{Liu et~al.}{2021}]{SwinTransformer}
\begin{botherref}
\oauthor{\bsnm{Liu}, \binits{Z.}},
\oauthor{\bsnm{Lin}, \binits{Y.}},
\oauthor{\bsnm{Cao}, \binits{Y.}},
\oauthor{\bsnm{Hu}, \binits{H.}},
\oauthor{\bsnm{Wei}, \binits{Y.}},
\oauthor{\bsnm{Zhang}, \binits{Z.}},
\oauthor{\bsnm{Lin}, \binits{S.}},
\oauthor{\bsnm{Guo}, \binits{B.}}:
Swin transformer: Hierarchical vision transformer using shifted windows.
CoRR
\textbf{abs/2103.14030}
(2021)
{\href{https://arxiv.org/abs/2103.14030}{{2103.14030}}}
\end{botherref}
\endbibitem

%%% 16
\bibitem[\protect\citeauthoryear{Ren et~al.}{2019}]{HRNet}
\begin{barticle}
\bauthor{\bsnm{Ren}, \binits{Z.}},
\bauthor{\bsnm{Xu}, \binits{Z.}},
\bauthor{\bsnm{Lam}, \binits{E.Y.M.}}:
\batitle{{End-to-end deep learning framework for digital holographic reconstruction}}.
\bjtitle{Advanced Photonics}
\bvolume{1}(\bissue{1}),
\bfpage{016004}
(\byear{2019})
\doiurl{10.1117/1.AP.1.1.016004}
\end{barticle}
\endbibitem

%%% 17
\bibitem[\protect\citeauthoryear{Zhang et~al.}{2018}]{Zhang:18}
\begin{barticle}
\bauthor{\bsnm{Zhang}, \binits{G.}},
\bauthor{\bsnm{Guan}, \binits{T.}},
\bauthor{\bsnm{Shen}, \binits{Z.}},
\bauthor{\bsnm{Wang}, \binits{X.}},
\bauthor{\bsnm{Hu}, \binits{T.}},
\bauthor{\bsnm{Wang}, \binits{D.}},
\bauthor{\bsnm{He}, \binits{Y.}},
\bauthor{\bsnm{Xie}, \binits{N.}}:
\batitle{Fast phase retrieval in off-axis digital holographic microscopy through deep learning}.
\bjtitle{Opt. Express}
\bvolume{26}(\bissue{15}),
\bfpage{19388}--\blpage{19405}
(\byear{2018})
\doiurl{10.1364/OE.26.019388}
\end{barticle}
\endbibitem

%%% 18
\bibitem[\protect\citeauthoryear{Li et~al.}{2022}]{app122010656}
\begin{botherref}
\oauthor{\bsnm{Li}, \binits{Z.}},
\oauthor{\bsnm{Chen}, \binits{Y.}},
\oauthor{\bsnm{Sun}, \binits{J.}},
\oauthor{\bsnm{Jin}, \binits{Y.}},
\oauthor{\bsnm{Shen}, \binits{Q.}},
\oauthor{\bsnm{Gao}, \binits{P.}},
\oauthor{\bsnm{Chen}, \binits{Q.}},
\oauthor{\bsnm{Zuo}, \binits{C.}}:
High bandwidth-utilization digital holographic reconstruction using an untrained neural network.
Applied Sciences
\textbf{12}(20)
(2022)
\doiurl{10.3390/app122010656}
\end{botherref}
\endbibitem

%%% 19
\bibitem[\protect\citeauthoryear{Wang et~al.}{2019}]{Y-net}
\begin{barticle}
\bauthor{\bsnm{Wang}, \binits{K.}},
\bauthor{\bsnm{Dou}, \binits{J.}},
\bauthor{\bsnm{Kemao}, \binits{Q.}},
\bauthor{\bsnm{Di}, \binits{J.}},
\bauthor{\bsnm{Zhao}, \binits{J.}}:
\batitle{Y-net: a one-to-two deep learning framework for digital holographic reconstruction}.
\bjtitle{Opt. Lett.}
\bvolume{44}(\bissue{19}),
\bfpage{4765}--\blpage{4768}
(\byear{2019})
\doiurl{10.1364/OL.44.004765}
\end{barticle}
\endbibitem

%%% 20
\bibitem[\protect\citeauthoryear{Cuche et~al.}{2000}]{Cuche:00}
\begin{barticle}
\bauthor{\bsnm{Cuche}, \binits{E.}},
\bauthor{\bsnm{Marquet}, \binits{P.}},
\bauthor{\bsnm{Depeursinge}, \binits{C.}}:
\batitle{Spatial filtering for zero-order and twin-image elimination in digital off-axis holography}.
\bjtitle{Appl. Opt.}
\bvolume{39}(\bissue{23}),
\bfpage{4070}--\blpage{4075}
(\byear{2000})
\doiurl{10.1364/AO.39.004070}
\end{barticle}
\endbibitem

%%% 21
\bibitem[\protect\citeauthoryear{Mann et~al.}{2005}]{mann2005high}
\begin{barticle}
\bauthor{\bsnm{Mann}, \binits{C.}},
\bauthor{\bsnm{Yu}, \binits{L.}},
\bauthor{\bsnm{Lo}, \binits{C.-M.}},
\bauthor{\bsnm{Kim}, \binits{M.}}:
\batitle{High-resolution quantitative phase-contrast microscopy by digital holography}.
\bjtitle{Optics express}
\bvolume{13}(\bissue{22}),
\bfpage{8693}--\blpage{8698}
(\byear{2005})
\doiurl{10.1364/opex.13.008693}
\end{barticle}
\endbibitem

%%% 22
\bibitem[\protect\citeauthoryear{Matrecano et~al.}{2015}]{Butterworth_Filter}
\begin{barticle}
\bauthor{\bsnm{Matrecano}, \binits{M.}},
\bauthor{\bsnm{Memmolo}, \binits{P.}},
\bauthor{\bsnm{Miccio}, \binits{L.}},
\bauthor{\bsnm{Persano}, \binits{A.}},
\bauthor{\bsnm{Quaranta}, \binits{F.}},
\bauthor{\bsnm{Siciliano}, \binits{P.}},
\bauthor{\bsnm{Ferraro}, \binits{P.}}:
\batitle{Improving holographic reconstruction by automatic butterworth filtering for microelectromechanical systems characterization}.
\bjtitle{Applied optics}
\bvolume{54}(\bissue{11}),
\bfpage{3428}--\blpage{3432}
(\byear{2015})
\doiurl{10.1364/AO.54.003428}
\end{barticle}
\endbibitem

%%% 23
\bibitem[\protect\citeauthoryear{Weng et~al.}{2014}]{WENG20142633}
\begin{barticle}
\bauthor{\bsnm{Weng}, \binits{J.}},
\bauthor{\bsnm{Li}, \binits{H.}},
\bauthor{\bsnm{Zhang}, \binits{Z.}},
\bauthor{\bsnm{Zhong}, \binits{J.}}:
\batitle{Design of adaptive spatial filter at uniform standard for automatic analysis of digital holographic microscopy}.
\bjtitle{Optik}
\bvolume{125}(\bissue{11}),
\bfpage{2633}--\blpage{2637}
(\byear{2014})
\doiurl{10.1016/j.ijleo.2013.11.035}
\end{barticle}
\endbibitem

%%% 24
\bibitem[\protect\citeauthoryear{Hong et~al.}{2017}]{WeightedAdaptiveSpatialFilter}
\begin{barticle}
\bauthor{\bsnm{Hong}, \binits{Y.}},
\bauthor{\bsnm{Shi}, \binits{T.}},
\bauthor{\bsnm{Wang}, \binits{X.}},
\bauthor{\bsnm{Zhang}, \binits{Y.}},
\bauthor{\bsnm{Chen}, \binits{K.}},
\bauthor{\bsnm{Liao}, \binits{G.}}:
\batitle{Weighted adaptive spatial filtering in digital holographic microscopy}.
\bjtitle{Optics Communications}
\bvolume{382},
\bfpage{624}--\blpage{631}
(\byear{2017})
\doiurl{10.1016/j.optcom.2016.08.056}
\end{barticle}
\endbibitem

%%% 25
\bibitem[\protect\citeauthoryear{He et~al.}{2016}]{he2016automated}
\begin{barticle}
\bauthor{\bsnm{He}, \binits{X.}},
\bauthor{\bsnm{Nguyen}, \binits{C.V.}},
\bauthor{\bsnm{Pratap}, \binits{M.}},
\bauthor{\bsnm{Zheng}, \binits{Y.}},
\bauthor{\bsnm{Wang}, \binits{Y.}},
\bauthor{\bsnm{Nisbet}, \binits{D.R.}},
\bauthor{\bsnm{Williams}, \binits{R.J.}},
\bauthor{\bsnm{Rug}, \binits{M.}},
\bauthor{\bsnm{Maier}, \binits{A.G.}},
\bauthor{\bsnm{Lee}, \binits{W.M.}}:
\batitle{Automated fourier space region-recognition filtering for off-axis digital holographic microscopy}.
\bjtitle{Biomed. Opt. Express}
\bvolume{7}(\bissue{8}),
\bfpage{3111}--\blpage{3123}
(\byear{2016})
\doiurl{10.1364/BOE.7.003111}
\end{barticle}
\endbibitem

%%% 26
\bibitem[\protect\citeauthoryear{Xiao et~al.}{2019}]{CNNFilter}
\begin{barticle}
\bauthor{\bsnm{Xiao}, \binits{W.}},
\bauthor{\bsnm{Wang}, \binits{Q.}},
\bauthor{\bsnm{Pan}, \binits{F.}},
\bauthor{\bsnm{Cao}, \binits{R.}},
\bauthor{\bsnm{Wu}, \binits{X.}},
\bauthor{\bsnm{Sun}, \binits{L.}}:
\batitle{Adaptive frequency filtering based on convolutional neural networks in off-axis digital holographic microscopy}.
\bjtitle{Biomed Opt Express}
\bvolume{10}(\bissue{4}),
\bfpage{1613}--\blpage{1626}
(\byear{2019})
\doiurl{10.1364/BOE.10.001613}
\end{barticle}
\endbibitem

%%% 27
\bibitem[\protect\citeauthoryear{{Ovizio Imaging Systems}}{}]{osone}
\begin{botherref}
\oauthor{\bsnm{{Ovizio Imaging Systems}}}:
OsOne Software.
[Online]. Available: \url{https://ovizio.com/software-osone/}
\end{botherref}
\endbibitem

%%% 28
\bibitem[\protect\citeauthoryear{Ronneberger et~al.}{2015}]{U-Net}
\begin{botherref}
\oauthor{\bsnm{Ronneberger}, \binits{O.}},
\oauthor{\bsnm{Fischer}, \binits{P.}},
\oauthor{\bsnm{Brox}, \binits{T.}}:
U-net: Convolutional networks for biomedical image segmentation.
CoRR
\textbf{abs/1505.04597}
(2015)
{\href{https://arxiv.org/abs/1505.04597}{{1505.04597}}}
\end{botherref}
\endbibitem

%%% 29
\bibitem[\protect\citeauthoryear{He et~al.}{2015}]{ResNet}
\begin{botherref}
\oauthor{\bsnm{He}, \binits{K.}},
\oauthor{\bsnm{Zhang}, \binits{X.}},
\oauthor{\bsnm{Ren}, \binits{S.}},
\oauthor{\bsnm{Sun}, \binits{J.}}:
Deep residual learning for image recognition.
CoRR
\textbf{abs/1512.03385}
(2015)
{\href{https://arxiv.org/abs/1512.03385}{{1512.03385}}}
\end{botherref}
\endbibitem

%%% 30
\bibitem[\protect\citeauthoryear{R{\"o}hrl et~al.}{2023}]{rohrl2023explainable}
\begin{bchapter}
\bauthor{\bsnm{R{\"o}hrl}, \binits{S.}},
\bauthor{\bsnm{Bernhard}, \binits{L.}},
\bauthor{\bsnm{Lengl}, \binits{M.}},
\bauthor{\bsnm{Klenk}, \binits{C.}},
\bauthor{\bsnm{Heim}, \binits{D.}},
\bauthor{\bsnm{Knopp}, \binits{M.}},
\bauthor{\bsnm{Schumann}, \binits{S.}},
\bauthor{\bsnm{Hayden}, \binits{O.}},
\bauthor{\bsnm{Diepold}, \binits{K.}}:
\bctitle{Explainable feature learning with variational autoencoders for holographic image analysis.}
In: \bbtitle{BIOIMAGING},
pp. \bfpage{69}--\blpage{77}
(\byear{2023})
\end{bchapter}
\endbibitem

%%% 31
\bibitem[\protect\citeauthoryear{Verrier and Atlan}{2011}]{Verrier:11}
\begin{barticle}
\bauthor{\bsnm{Verrier}, \binits{N.}},
\bauthor{\bsnm{Atlan}, \binits{M.}}:
\batitle{Off-axis digital hologram reconstruction: some practical considerations}.
\bjtitle{Appl. Opt.}
\bvolume{50}(\bissue{34}),
\bfpage{136}--\blpage{146}
(\byear{2011})
\doiurl{10.1364/AO.50.00H136}
\end{barticle}
\endbibitem

%%% 32
\bibitem[\protect\citeauthoryear{Dubois and Yourassowsky}{2004}]{patent1}
\begin{botherref}
\oauthor{\bsnm{Dubois}, \binits{F.}},
\oauthor{\bsnm{Yourassowsky}, \binits{C.}}:
US Patent 7,362,449 B2.
7,362,449 B2,
2004.
p. 1
\end{botherref}
\endbibitem

%%% 33
\bibitem[\protect\citeauthoryear{Dubois and Yourassowsky}{2015}]{patent2}
\begin{botherref}
\oauthor{\bsnm{Dubois}, \binits{F.}},
\oauthor{\bsnm{Yourassowsky}, \binits{C.}}:
US Patent 9,207,638 B2.
9,207,638 B2,
2015.
p. 1
\end{botherref}
\endbibitem

%%% 34
\bibitem[\protect\citeauthoryear{Kanopoulos et~al.}{1988}]{Sobel}
\begin{barticle}
\bauthor{\bsnm{Kanopoulos}, \binits{N.}},
\bauthor{\bsnm{Vasanthavada}, \binits{N.}},
\bauthor{\bsnm{Baker}, \binits{R.L.}}:
\batitle{Design of an image edge detection filter using the sobel operator}.
\bjtitle{IEEE Journal of solid-state circuits}
\bvolume{23}(\bissue{2}),
\bfpage{358}--\blpage{367}
(\byear{1988})
\end{barticle}
\endbibitem

%%% 35
\bibitem[\protect\citeauthoryear{Kingma and Ba}{2017}]{Adam}
\begin{botherref}
\oauthor{\bsnm{Kingma}, \binits{D.P.}},
\oauthor{\bsnm{Ba}, \binits{J.}}:
Adam: A Method for Stochastic Optimization
(2017).
\url{https://arxiv.org/abs/1412.6980}
\end{botherref}
\endbibitem

%%% 36
\bibitem[\protect\citeauthoryear{Paszke et~al.}{2019}]{Pytorch}
\begin{barticle}
\bauthor{\bsnm{Paszke}, \binits{A.}},
\bauthor{\bsnm{Gross}, \binits{S.}},
\bauthor{\bsnm{Massa}, \binits{F.}},
\bauthor{\bsnm{Lerer}, \binits{A.}},
\bauthor{\bsnm{Bradbury}, \binits{J.}},
\bauthor{\bsnm{Chanan}, \binits{G.}},
\bauthor{\bsnm{Killeen}, \binits{T.}},
\bauthor{\bsnm{Lin}, \binits{Z.}},
\bauthor{\bsnm{Gimelshein}, \binits{N.}},
\bauthor{\bsnm{Antiga}, \binits{L.}},
\bauthor{\bsnm{Desmaison}, \binits{A.}},
\bauthor{\bsnm{Kopf}, \binits{A.}},
\bauthor{\bsnm{Yang}, \binits{E.}},
\bauthor{\bsnm{DeVito}, \binits{Z.}},
\bauthor{\bsnm{Raison}, \binits{M.}},
\bauthor{\bsnm{Tejani}, \binits{A.}},
\bauthor{\bsnm{Chilamkurthy}, \binits{S.}},
\bauthor{\bsnm{Steiner}, \binits{B.}},
\bauthor{\bsnm{Fang}, \binits{L.}},
\bauthor{\bsnm{Bai}, \binits{J.}},
\bauthor{\bsnm{Chintala}, \binits{S.}}:
\batitle{Pytorch: An imperative style, high-performance deep learning library}.
\bjtitle{Advances in Neural Information Processing Systems}
\bvolume{32},
\bfpage{8024}--\blpage{8035}
(\byear{2019})
\end{barticle}
\endbibitem

%%% 37
\bibitem[\protect\citeauthoryear{van~der Walt et~al.}{2014}]{scikit-image}
\begin{barticle}
\bauthor{\bsnm{Walt}, \binits{S.}},
\bauthor{\bsnm{{S}ch\"onberger}, \binits{J.L.}},
\bauthor{\bsnm{{Nunez-Iglesias}}, \binits{J.}},
\bauthor{\bsnm{{B}oulogne}, \binits{F.}},
\bauthor{\bsnm{{W}arner}, \binits{J.D.}},
\bauthor{\bsnm{{Y}ager}, \binits{N.}},
\bauthor{\bsnm{{G}ouillart}, \binits{E.}},
\bauthor{\bsnm{{Y}u}, \binits{T.}},
\bauthor{\bsnm{contributors}}:
\batitle{scikit-image: image processing in {P}ython}.
\bjtitle{PeerJ}
\bvolume{2},
\bfpage{453}
(\byear{2014})
\doiurl{10.7717/peerj.453}
\end{barticle}
\endbibitem

%%% 38
\bibitem[\protect\citeauthoryear{Girshovitz and Shaked}{2013}]{filter_refernce_beam}
\begin{barticle}
\bauthor{\bsnm{Girshovitz}, \binits{P.}},
\bauthor{\bsnm{Shaked}, \binits{N.T.}}:
\batitle{Compact and portable low-coherence interferometer with off-axis geometry for quantitative phase microscopy and nanoscopy}.
\bjtitle{Opt. Express}
\bvolume{21}(\bissue{5}),
\bfpage{5701}--\blpage{5714}
(\byear{2013})
\doiurl{10.1364/OE.21.005701}
\end{barticle}
\endbibitem

\end{thebibliography}

\newpage
\begin{appendices}

\section{Spatial Separation of Off-Axis Holograms in Frequency Domain} \label{sec: a}
\begin{figure}[htb]
    \centering
    \includegraphics[width=1\textwidth]{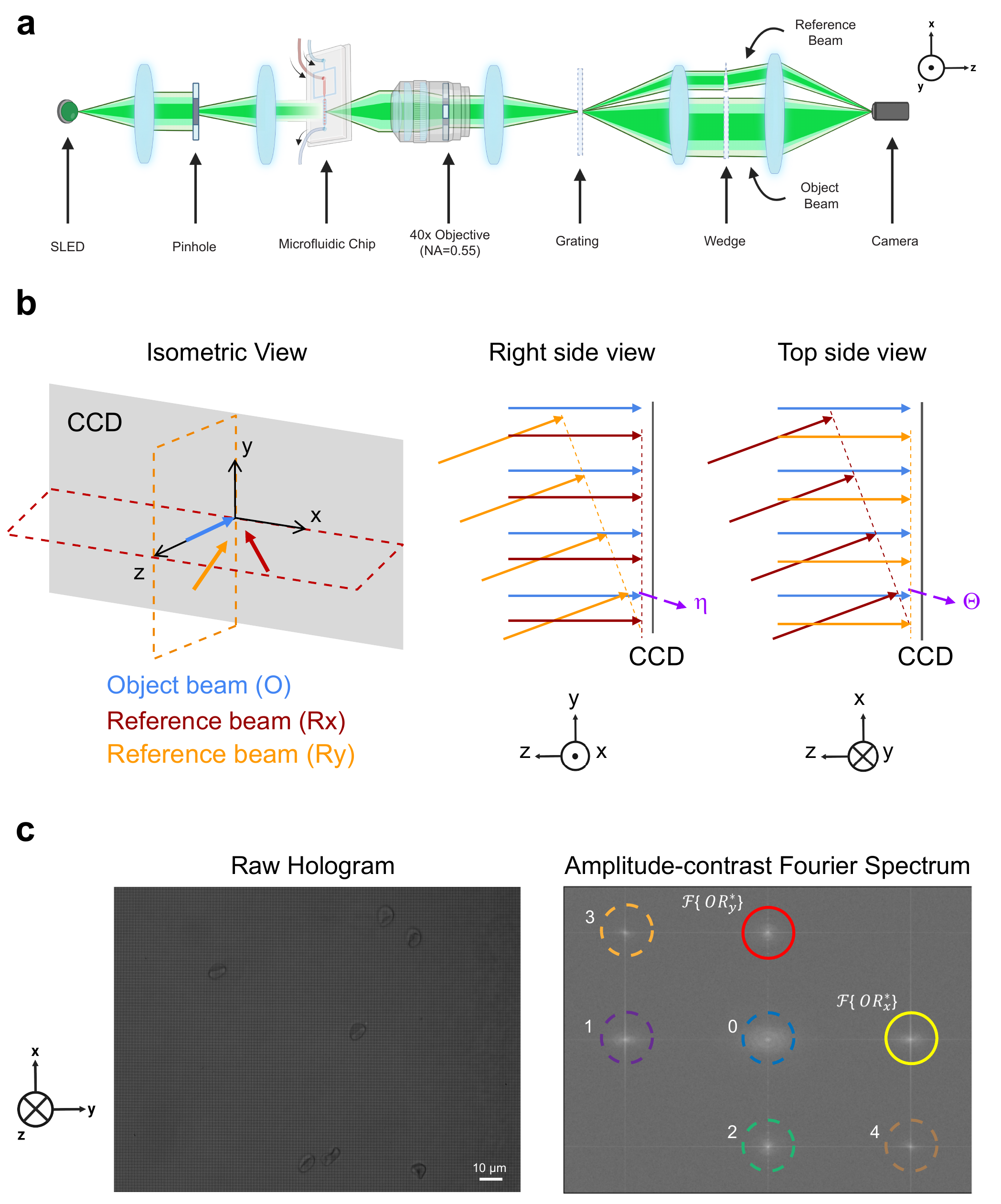}
    \caption{(a) Experimental setup of our DHM. Technical details remain undisclosed by the device manufacturer. The operating principle of DHM is as follows: A superluminescent light-emitting (SLED) provides partially coherent K\"oehler illumination, which is filtered through a pinhole. After the light scatters from blood cells flowing within a microfluidic channel, it is collected by a 40$\times$ objective lens with a numerical aperture (NA) of 0.55. A grating then splits the light, and a wedge introduces a slight tilt to the upper portion of the light beam. The reference and object beams are eventually superimposed and recorded by the imaging sensor to generate the holograms. (b) Illustration of the tilt angles between the object beam and the reference beams in our DHM setup. The two reference beams are titled along $x$ and $y$ axes, respectively. (c) An illustrative example of a raw hologram image and the corresponding spectrum after Fourier transform (denoted as $\mathcal{F}$), revealing the spatial separation of image signals: 0, $\mathcal{F}(R_xR_x^* + R_yR_y^* + OO^*)$; 1, $\mathcal{F}(O^*R_x)$; 2, $\mathcal{F}(O^*R_y)$; 3, $\mathcal{F}(R_xR_y^*)$; 4, $\mathcal{F}(R_x^*R_y)$. More details are provided in the Method and Supplementary.}
    \label{fig: s1}
\end{figure}

Based on the Fourier spectrum, we deduce that there are two reference beams, as illustrated in Figure. \ref{fig: s1}. These two reference beams were tilted along the $x$ and $y$ axes, corresponding waves denoted as $R_x(x,y)$ and $R_y(x,y)$, respectively. The CCD sensor records the intensity distribution $I_H(x, y)$, representing the interference pattern of the object beam wave $O(x,y)$ and two reference beam waves:
\begin{align}
    I_H & = (R_x + R_y + O)(R_x + R_y + O)^* \nonumber \\
    & = R_xR_x^* + R_yR_y^* + OO^* + OR_x^* + R_xO^* + OR_y^* + R_yO^* + R_xR_y^* + R_x^*R_y \label{eq: s1}
\end{align}
where $^*$ denotes the complex conjugation. When $I_H$ is recorded with sample, denoted as $I_{H,s}$, the corresponding object wave is:
\begin{equation}
        O_s(x, y) = A_{O,s}(x, y)e^{i \phi_O} e^{i \Delta \phi(x, y)} \label{eq: s2} \\
\end{equation}
where $A_{O, s}$ is the amplitude of the object wave, $\phi_O$ is the object wave phase, and $\Delta \phi(x, y)$ is the phase shift caused by the refractive index variations of the sample. When $I_H$ is recorded in the absence of sample, denoted as $I_{H,b}$, the corresponding object wave is:
\begin{equation}
    O_b(x, y) = A_{O,b}(x, y)e^{i \phi_O} \label{eq: s3} \\
\end{equation}
No sample-induced phase shift is present in $O_b(x,y)$.

The sample signal is filtered from the reference waves \cite{filter_refernce_beam}. Therefore, they remains the same in both $I_{H,s}$ and $I_{H,b}$:
\begin{align}
    R_{x,s}(x, y) = R_{x,b}(x,y) & = A_{Rx}e^{i \phi_{Rx}} e^{-i k x \sin \theta} \label{eq: s4} \\
    R_{y,s}(x, y) = R_{y,b}(x,y) & = A_{Ry}e^{i \phi_{Ry}} e^{-i k y \sin \eta} \label{eq: s5} \\
    k & = 2\pi / \lambda
\end{align}
where $\lambda$ is the laser wavelength; $A_{Rx}$ and $A_{Ry}$ are the amplitudes of the reference waves that are uniform over the CCD. The phase shift $-k x \sin \theta$ is caused by the tilted angle $\theta$ between the phase plane of $R_x$ and the CCD, as similarly for $-k x \sin \eta$ and $\eta$. Substituting Eq. \ref{eq: s2}-\ref{eq: s5} into Eq. \ref{eq: s1}, we have:
\begin{align}
    I_{H,s}(x,y) = & A_{Rx}^2(x,y) + A_{Ry}^2(x,y) + A_O^2(x,y) \nonumber \\
    & + A_{O,s}(x,y)A_{Rx}e^{i[\phi_O - \phi_{Rx} + \Delta \phi(x, y) + k x \sin \theta]} \nonumber \\
    & + A_{O,s}(x,y)A_{Rx}e^{i[\phi_{Rx} - \phi_O - \Delta \phi(x, y) - k x \sin \theta]} \nonumber \\
    & + A_{O,s}(x,y)A_{Ry}e^{i[\phi_O - \phi_{Ry} + i \Delta \phi(x, y) + k y \sin \eta]} \nonumber \\
    & + A_{O,s}(x,y)A_{Ry}e^{i[\phi_{Ry} - \phi_O -\Delta \phi(x, y) - k y \sin \eta]} \nonumber \\
    & + A_{Ry}A_{Rx}e^{i(\phi_{Rx} - \phi_{Ry} - k x \sin \theta + ky \sin \eta)} \nonumber \\
    & + A_{Rx}A_{Ry}e^{i(\phi_{Ry} - \phi_{Rx} + k y \sin \eta - kx \sin \theta)} \nonumber \\
    \label{eq: s6}
\end{align}

After performing Fourier transformation (denoted as $\mathcal{F}$), these terms become:
\begin{align}
    & \mathcal{F}(R_xR_x^* + R_yR_y^* + OO^*) = \iint [A_{Rx}^2 + A_{Ry}^2 + A_O(x,y)^2]e^{-2i\pi (ux + vy)} dx dy  \\
    & \mathcal{F}(OR_x^*) = e^{i(\phi_O - \phi_{Rx})} \iint A_{O,s}(x,y)A_{Rx}e^{i \Delta \phi(x, y)}e^{-2i \pi [(u - \sin \theta / \lambda) x + vy ]} dx dy \label{eq: s7} \\
    & \mathcal{F}(R_xO^*) = e^{i(\phi_{Rx} - \phi_O)} \iint A_{O,s}(x,y)A_{Rx}e^{-i \Delta \phi(x, y)}e^{-2i \pi [(u + \sin \theta / \lambda) x + vy ]} dx dy \\
    & \mathcal{F}(OR_y^*) = e^{i(\phi_O - \phi_{Ry})} \iint
    A_{O,s}(x,y)A_{Ry}e^{i \Delta \phi(x, y)}e^{-2i \pi [ux + (v - \sin \eta / \lambda) y ]} dx dy \\
    & \mathcal{F}(R_yO^*) = e^{i(\phi_{Ry} - \phi_O)} \iint A_{O,s}(x,y)A_{Ry}e^{-i \Delta \phi(x, y)}e^{-2i \pi [ux + (v + \sin \eta / \lambda) y ]} dx dy \\
    & \mathcal{F}(R_xR_y^*) = e^{i(\phi_{Rx} - \phi_{Ry})} \iint A_{Rx}A_{Ry}e^{-2i \pi [(u + \sin \theta / \lambda) x + (v - \sin \eta / \lambda) y ]} dx dy \\
    & \mathcal{F}(R_yR_x^*) = e^{i(\phi_{Ry} - \phi_{Rx})} \iint
    A_{Rx}A_{Ry}e^{-2i \pi [(u - \sin \theta / \lambda) x + (v + \sin \eta / \lambda) y ]} dx dy
\end{align}
Eq. \ref{eq: s7} shows that $\mathcal{F}(OR_x^*)$ are shifted by $\sin \theta / \lambda$ along the $x$ axis from the origin. Similarly, all the other terms are shifted along different directions in the Fourier spectrum except the zero-order term $\mathcal{F}(R_xR_x^* + R_yR_y^* + OO^*)$, as illustrated in Fig. 1b.

\newpage
\section{Spatial filtering and demodulation} \label{sec: b}
Using the spatial separation in the Fourier spectrum, the real image (either $OR^*_x$ or $OR_y^*$) can be extracted directly from the hologram. Their frequency shift caused by the tilt angle ($\sin \theta / \lambda$ or $\sin \eta / \lambda$) could be corrected by shifting them back to the center of the spectrum. The detailed procedure is shown in Fig. \ref{fig: s2}. After the operations in the Fourier domain and the inverse Fourier transformation, the real images become:
\begin{align}
    \Psi_s(x,y) &= A_{O,s}(x,y)A_R e^{i(\phi_O - \phi_R)}e^{i \Delta \phi(x, y)}   \label{eq: s8} \\
    \Psi_b(x,y) &= A_{O,b}(x,y)I_R e^{i(\phi_O - \phi_R)}  \label{eq: s9}
\end{align}
where $\Psi_s(x,y)$ and $\Psi_b(x,y)$ are derived from $I_{H,s}$ and $I_{H,b}$, respectively. $\Psi_s(x,y)$ could be further demodulated by dividing it by $\Psi_b(x,y)$:
\begin{equation}
    \Psi'(x,y) = \frac{\Psi_s(x,y)}{\Psi_b(x,y)} = \frac{A_{O,s}(x,y)}{A_{O,b}(x,y)}e^{i \Delta \phi(x, y)}   \label{eq: s10}
\end{equation}
where $\Delta \phi(x, y)$ provides the 3D information of the sample, and $A_{O,s}(x,y) / A_{O,b}(x,y)$ denotes the relative intensity changes induced by the sample.

\newpage
\section{Matrix multiplications with  Fourier Imager Heads} \label{sec: c}
\begin{figure}[htb]
    \centering
    \includegraphics[width=1\textwidth]{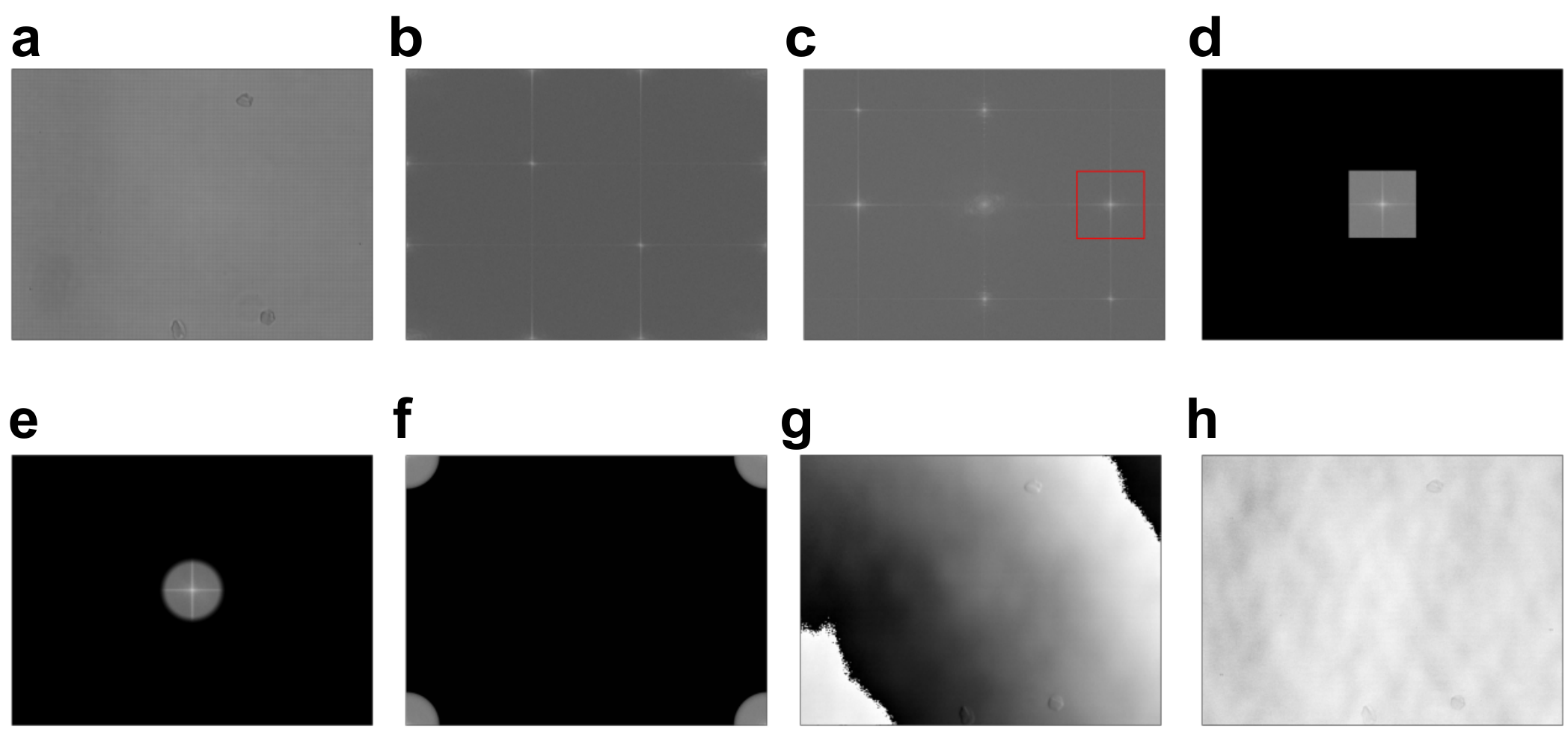}
    \caption{The classic hologram reconstruction procedure involving the Fourier frequency filtration with a circular filter. All the frequency spectra are illustrated by taking the log of amplitude. (a) Original hologram; (b) Frequency spectrum generated by 2D Fourier transform; (c) Shifting the origin of the frequency spectrum to the center. The target frequency domain is indicated in the red square; (d) Cropping the red square and shifting it to the center to remove the phase shift caused by the tilted reference beam; (e) Further filter the spectrum by pixel-wise multiplication with a mask matrix $M_{mask}$; (f) Shifting the origin of frequency spectrum back to the corners; (g) Phase image after inverse 2D Fourier transform from (f); (h) Amplitude image after inverse 2D Fourier transform from (f).}
    \label{fig: s2}
\end{figure}

As demonstrated in Table \ref{tbl: S1}, each step of the spatial filtering process in Fig. \ref{fig: s2} can be converted into a matrix multiplication, which could be further simplified as:
\begin{equation}
    \Tilde{M}_l = M^2_l M^1_l M^0_l \qquad
    \Tilde{M}_r = M^0_r M^1_r M^2_r \qquad
    \Tilde{M}_{mask} = M^2_l M_{mask} M^2_r \label{eq: s11}
\end{equation}
Clearly, $\Tilde{M}_l$, $\Tilde{M}_r$ and $\Tilde{M}_{mask}$ can be determined manually, then loaded into the neural network as initial weights.

\begin{table}[htb]
    \centering
    \begin{tabular}{ | c | c | c | }
        \hline
        \thead{Operation} & \thead{Step \\ in Fig. \ref{fig: s2}} & \thead{ Implementation by \\ matrix multiplication}\\
        \hline
        \makecell{Converting the origin of \\ frequency domain from image \\ corners to the center} & \makecell{\ref{fig: s2}b $\to$ \ref{fig: s2}c} & \makecell{$\mathcal{F}(I_H) \to M^0_l\mathcal{F}(I_H)M^0_r$} \\ 
        \hline
        \makecell{Cropping out the target \\ frequency region and shifting \\ it to the center} & \makecell{\ref{fig: s2}c $\to$ \ref{fig: s2}d} & \makecell{$M^0_l\mathcal{F}(I_H)M^0_r \to$ \\ $M^1_l M^0_l \mathcal{F}(I_H) M^0_r M^1_r$} \\ 
        \hline
        \makecell{Filtering out high frequencies} & \makecell{\ref{fig: s2}d $\to$ \ref{fig: s2}e} & \makecell{$ M^1_lM^0_l\mathcal{F}(I_H)M^0_rM^1_r \to $ \\ $ M^1_lM^0_l\mathcal{F}(I_H)M^0_rM^1_r \odot M_{mask}$} \\ 
        \hline
        \makecell{Converting the origin of \\ frequency domain from image \\ center to the corners} & \makecell{\ref{fig: s2}e $\to$ \ref{fig: s2}f} & \makecell{$M^1_lM^0_l\mathcal{F}(I_H)M^0_rM^1_r \odot M_{mask} \to$ \\ $M^2_l[M^1_lM^0_l\mathcal{F}(I_H)M^0_rM^1_r \odot M_{mask}]M^2_r$ } \\ 
      \hline
    \end{tabular}
    \caption{Detailed steps of spatial filtering in the frequency domain and corresponding implementations by matrix multiplication. $\odot$ denotes the element-wise matrix multiplication. All matrices, except the frequency domain image $F$, can be determined manually.}
    \label{tbl: S1}
\end{table}

\begin{figure}[htb]
    \centering
    \includegraphics[width=1\textwidth]{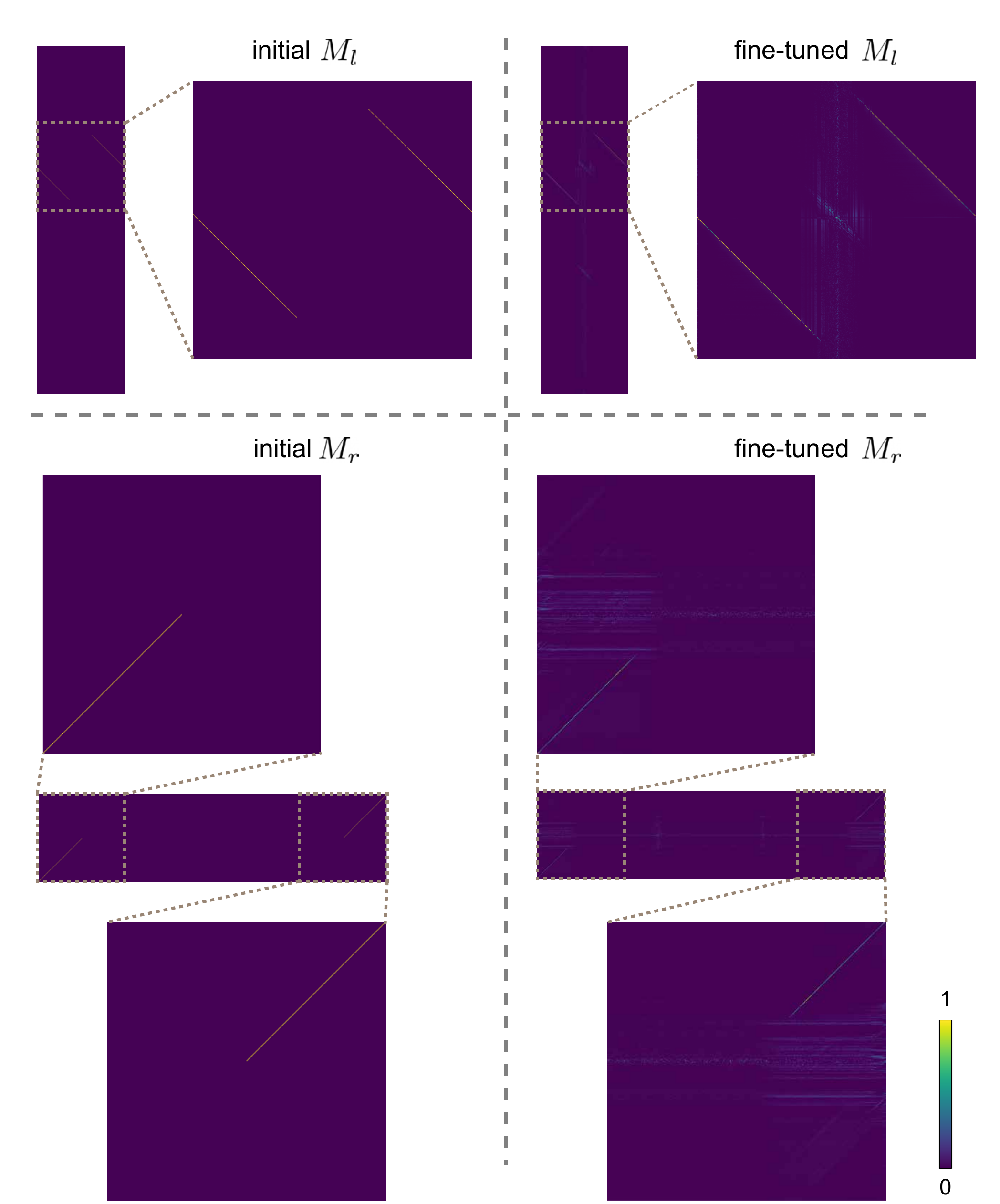}
    \caption{The initial and fine-tuned $M_l$ and $M_r$ in the first FIH.}
    \label{fig: s3}
\end{figure}

\begin{figure}[htb]
    \centering
    \includegraphics[width=1\textwidth]{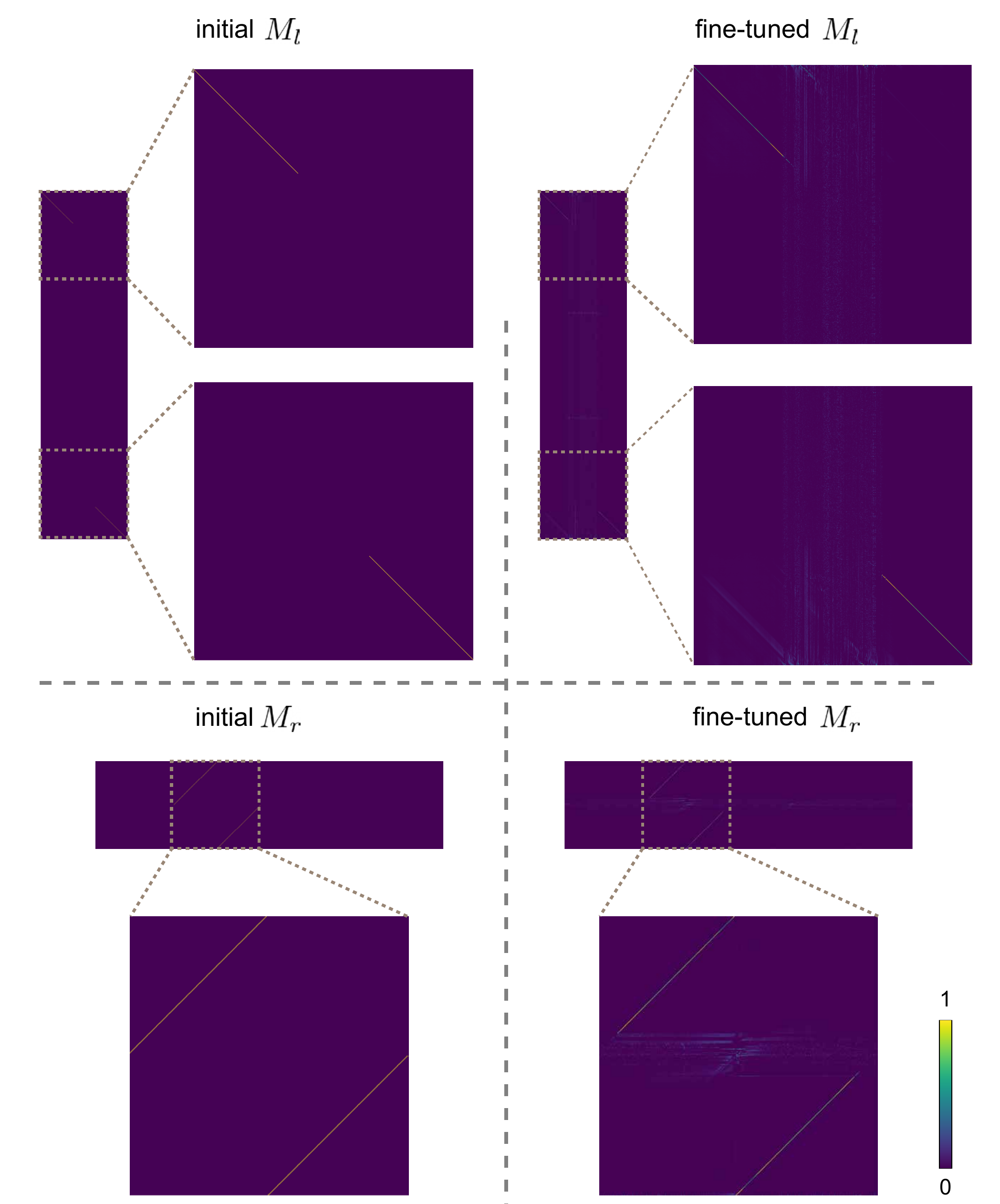}
    \caption{The initial and fine-tuned $M_l$ and $M_r$ in the second FIH.}
    \label{fig: s4}
\end{figure}

\begin{table}[htb]
    \centering
    \begin{tabular}{ | c | c | c | c | c |}
        \hline
        \thead{Matrix} & \thead{\makecell{Number of Weights\\(Total)}} & \thead{\makecell{Number of Weights\\($<-0.1$)}} & \thead{\makecell{Number of Weights\\($>1$)}}  & \thead{\makecell{Percentage of Weights\\($\in[-0.1, 1]$)}}\\
        \hline
        initial $M_l$ & 589,824 & 0 & 0 & 100\% \\ 
        \hline
        $M_l$ (FIH 1) & 589,824 & 3,469 & 5 & 99.4\% \\ 
        \hline
        $M_l$ (FIH 2) & 589,824 & 1,469 & 13 & 99.7\% \\ 
        \hline
        initial $M_r$ & 1,048,576 & 0 & 0 & 100\% \\ 
        \hline
        $M_r$ (FIH 1) & 1,048,576 & 1,980 & 93 & 99.8\% \\ 
        \hline
        $M_r$ (FIH 2) & 1,048,576 & 4,123 & 37 & 99.6\% \\ 
        \hline
        initial $M_{mask}$ & 196,608 & 0 & 0 & 100\% \\ 
        \hline
        $M_{mask}$(FIH 1) & 196,608 & 1,122 & 90 & 99.3\% \\ 
        \hline
        $M_{mask}$(FIH 2) & 196,608 & 792 & 14 & 99.6\% \\ 
        \hline

      \hline
    \end{tabular}
    \caption{Statistics of the value range of $M_l$, $M_r$, and $M_{mask}$. Their initial weights consist only of 0 or 1. No constraint requires the weights of these matrices to remain within the [0, 1] range. However, after training, most values still fall within this range.}
    \label{tbl: S2}
\end{table}
\end{appendices}


\end{document}